\documentclass{emulateapj}

\usepackage{lscape}
\newcommand{\erg}{\mbox{$\rm\,erg$}}
\newcommand{\kev}{\mbox{$\rm\,keV$}}
\newcommand{\cm}{\mbox{$\rm\,cm$}}
\newcommand{\ev}{\mbox{$\rm\,eV$}}

\newcommand{\s}{\mbox{$\rm\,s$}}

\newcommand{\mpc}{\mbox{$\rm\,Mpc$}}
\newcommand{\kpc}{\mbox{$\rm\,kpc$}}

\newcommand{\beq}{\begin{equation}}
\newcommand{\eeq}{\end{equation}}

\newcommand{\LNVSS}{\ensuremath{L_{\rm NVSS}}}

\newcommand{\rhomass}{\ensuremath{\rho_{\rm 2MASS}}}

\usepackage{color}

\shorttitle{Morphological Evidence for AGN Feedback in Normal Ellipticals}
\shortauthors{Diehl \& Statler}

\begin{document}

\title{The Hot Interstellar Medium in Normal Elliptical Galaxies. II.\linebreak Morphological Evidence for AGN Feedback}

\author{Steven Diehl\altaffilmark{1,2} and Thomas S. Statler\altaffilmark{1}}
\altaffiltext{1}{Astrophysical Institute, Department of Physics and Astronomy,
251B Clippinger Research Laboratories, Ohio University, Athens, OH
45701, USA}
\altaffiltext{2}{Theoretical Astrophysics Group T-6, Mailstop B227, Los Alamos National Laboratory, P.O. Box 1663, Los Alamos, NM 87545, USA (present address)}
\email{diehl@lanl.gov, statler@ohio.edu}

\begin{abstract}
We report on the discovery of a new quantitative relationship between X-ray gas morphology and radio and X-ray AGN luminosities in normal elliptical galaxies. This is the second paper in a series using data on 54 objects from the {\it Chandra} public archive and builds on the findings of Paper I, which demonstrated that hydrostatic equilibrium in elliptical galaxies holds, at best, only approximately and that the shape of the X-ray isophotes is unrelated to the shape of the gravitational potential. Instead, the gas is almost always asymmetrically disturbed. In this paper, we quantify the amount of asymmetry and study its correlation with other galaxy properties. We also determine radio powers and derive X-ray
AGN luminosities for our galaxy sample. We find that the amount of asymmetry in the gas is correlated with both measures of AGN activity, in the sense that the hot gas is more disturbed in galaxies with higher radio and X-ray AGN luminosities. We find no evidence that galaxy density has significant effects on gas morphology. We do however find evidence for a correlation between gas asymmetry and the presence of hot ambient gas, which we interpret as a signature of hydrodynamic interactions with an external ambient medium. Surprisingly, the AGN--morphology connection persists all the way down to the weakest AGN luminosities in rather X-ray faint galaxies. This is strong morphological evidence that supports the general importance of AGN feedback, even in normal elliptical galaxies.
\end{abstract}



\section{Introduction}\label{s4.introduction}

In the first paper of this series \citep[][hereafter Paper
I]{DiehlGallery} we conducted a morphological analysis of the
hot interstellar medium (ISM) of normal elliptical galaxies, derived
from archived {\it Chandra} observations of 54 objects. This paper
introduced a new technique to isolate the diffuse hot gas emission
from the contaminating effects of unresolved point sources, and
presented a gallery of adaptively binned gas-only images. By fitting
elliptical isophotes to these images, we demonstrated that the
apparent flattening of the X-ray gas is completely uncorrelated with,
and often significantly larger than, that of the starlight. The absence of any
correlation, at radii inside the optical effective radius, implies that the hot gas cannot be in perfect hydrostatic
equilibrium, unless the potential is dominated by
a dark matter component completely unrelated to the stellar mass distribution. This is unlikely, since
within the effective
radius gravitational potentials are generally expected to be stellar-mass
dominated \citep[e.g.][]{MamonDarkmatterI,HumphreyDarkmatter}. Consequently, efforts to measure the shapes of dark
matter halos using X-ray gas isophotes may be fruitless in most cases
\citep[e.g.][]{BuoteGeomtest,BuoteNGC720}, and radial mass profiles
derived from X-ray data
\citep{HumphreyDarkmatter,FukazawaMassprofiles} could be in error by
factors of order unity.

This paper takes up the question left unaddressed by Paper I: what
{\it is} the origin of the gas flattening, if not gravitational? In
Paper I, we already excluded rotational flattening as the dominant
cause. A qualitative analysis of gas morphologies showed that
they are often asymmetric and disturbed. In this paper we will
quantify this asymmetry, argue that it and the isophotal ellipticity
reflect the same underlying disturbance, and show
strong evidence that the root causes are the
central active nucleus---even in those systems with
very weak active galactic nuclei (AGN)---and interactions with
ambient external gas.

Asymmetries are well known in X-ray studies of galaxies, groups, and
clusters. {\it Chandra} and {\it XMM-Newton} observations of clusters
of galaxies reveal strong deviations from smooth surface brightness
distributions. Many of these features are believed to be due to cold
fronts, caused by infalling substructure into the cluster center
\citep[e.g.][]{SarazinColdfront}, or gas sloshing in the gravitational
potential \citep[e.g.][]{MarkevitchSloshing}. In other cases,
depressions in the surface brightness distribution are found to be
coincident with extended radio emission
\citep[e.g.][]{McnamaraHydraA2}. These depressions, or ``bubbles'',
are generally believed to be pockets of low-density extremely hot
plasma, inflated by jets powered by the central AGN
\citep[see][for a recent review]{McNamaraAGNReview}. However, there are many cases where the
depressions have no obvious radio counterparts, which has brought them
the nickname ``ghost cavities'' \citep{McnamaraGhostCavities}. These
cavities are now interpreted to be relic bubbles that have detached from
the radio source and are buoyantly rising radially outward in the
cluster gas.

Unexpectedly, {\it Chandra} and {\it XMM} observations of normal elliptical galaxies have also revealed a wealth of highly disturbed gas morphologies with qualitatively similar properties to clusters. Many individual observations have been analyzed in detail and a variety of different explanations have been proposed. Sharp, one-sided drops in surface brightness are usually interpreted as signs of ram-pressure, distorting the gas in elliptical galaxies as they move through the ambient intracluster or intragroup medium, as in NGC~4472 \citep{IrwinNGC4472,BillerNGC4472} and NGC~1404 \citep{Machacek}. To explain the very asymmetric emission in NGC~7618, \citet{KraftNGC7618} even argue for a major group-group merger. Other observations are not interpreted as results of environmental effects, but rather as signatures of the central AGN. For a few cases, such as NGC~4374 \citep{Fin01} or NGC~4472 \citep{BillerNGC4472}, the association with the AGN is indicated clearly by the correspondence between the radio source and depressions in the X-ray gas distribution. While these cavities are generally surrounded by cool rims \citep[][and references therein]{McNamaraAGNReview}, two low-power radio galaxies (Centaurus~A and NGC~3801) have recently been identified with shocks surrounding the radio lobes \citep{KraftCentaurusA,KraftCentaurusA2,CrostonShocks}. In other objects, one can directly detect the X-ray counterpart of the radio jet \citep[e.g.][]{SambrunaXrayjets, HarrisXrayJet, HardcastleXrayJet}. However, most cases are less clear-cut. For example, the origin of the shock-like features seen in NGC~1553 \citep{BlantonNGC1553} or NGC~4636 \citep{JonesNGC4636} is still a mystery. The morphologies in other galaxies qualitatively resemble ghost cavities very close to the core \citep[e.g. NGC 5044,][]{BuoteNGC5044XMM}, without evidence for an active radio source.

These individual cases each come with their own analysis techniques and
interpretations, making general statements about elliptical
galaxies problematic. Detailed analyses are generally
limited to the X-ray brightest galaxies with clearly
identifiable features, which are often dismissed as special
cases. However, our
morphological survey of normal elliptical galaxies in
Paper I shows that disturbances and asymmetries in the gas are
actually the norm rather than the exception, even for relatively X-ray
faint galaxies. Most galaxies in the {\it Chandra\/} archive lack the signal, and
therefore the contrast, to reliably identify cavity or shock features individually. In this paper, we introduce a statistical
measure of asymmetry (the asymmetry index, $\eta$) that allows us to
treat all objects on an equal footing. We will show that the asymmetry
index is strongly correlated with two different measures of nuclear activity, and with
the presence of ambient hot gas.

The analysis in this paper begins with the elliptical isophotal fits
of Paper I, and is organized as follows. In \S\ref{s4.analysis} we
review the techniques of Paper I and describe the sample. We then
calculate smooth functional fits to the isophotal profiles, use these
fits to generate symmetric surface brightness models, and use the
residuals from these models to define the asymmetry index $\eta$ for
each galaxy. We show by simulations that $\eta$ is a monotonic measure
of the sort of asymmetry expected from multiple cavities, and that it
is independent of non-morphological parameters. We describe our
techniques to extract the central AGN X-ray luminosity and to
determine radio power and environment. In \S\ref{s4.results} we show
that the asymmetry and isophotal ellipticity are correlated, and thus probably
measure the same underlying disturbance. We show that neither of these
quantities depends on neighbor galaxy density, but that they are correlated with
measures of AGN power and the presence of ambient gas.
The consequences for AGN feedback in
galaxies and interactions with the ambient medium
are discussed in \S\ref{s4.discussion}, before we conclude
with a reiteration of our major results in \S\ref{s4.conclusions}.


\section{Data Analysis}\label{s4.analysis}

\begin{figure*}
\includegraphics[width=0.5\textwidth]{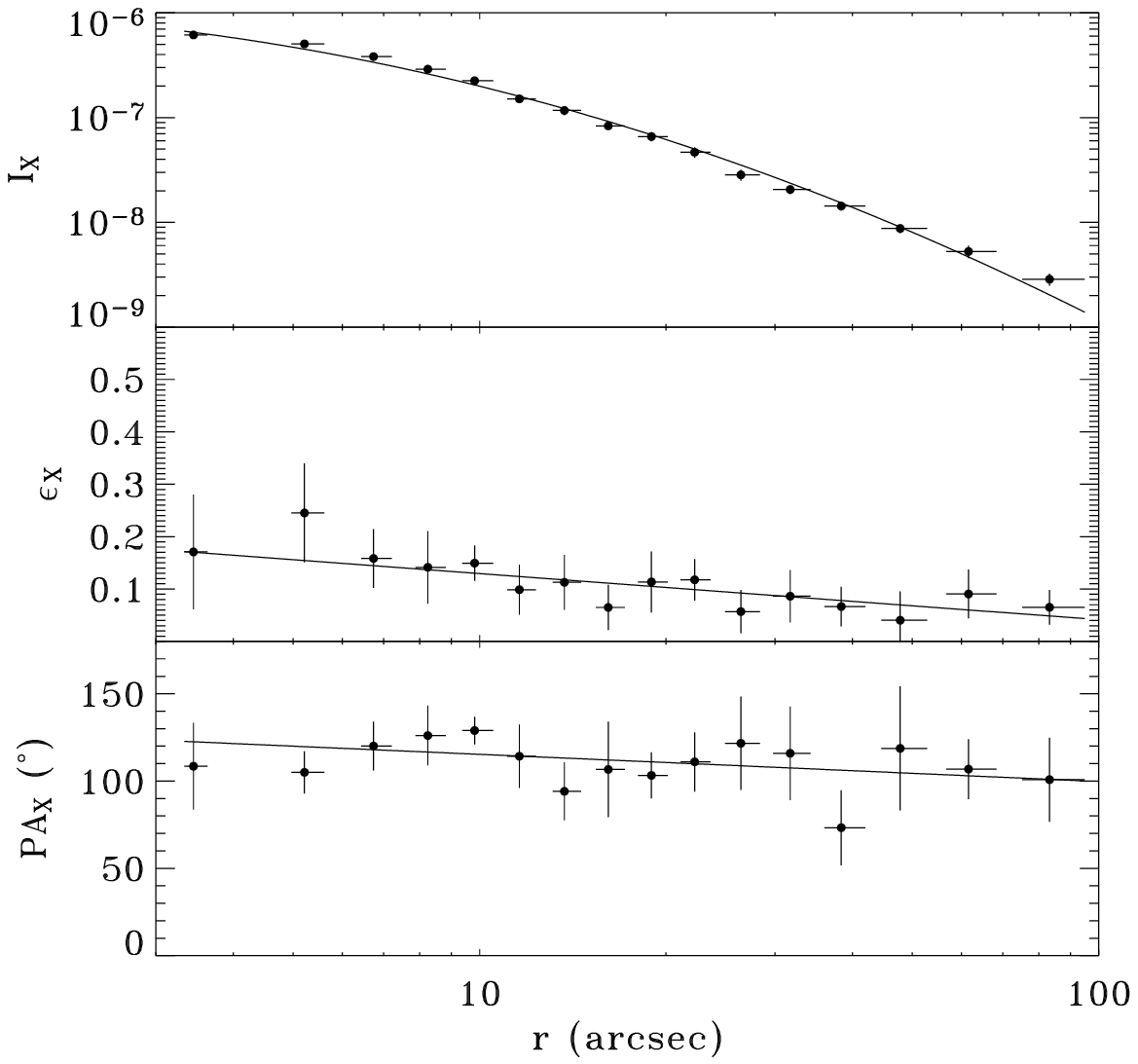}
\includegraphics[width=0.5\textwidth]{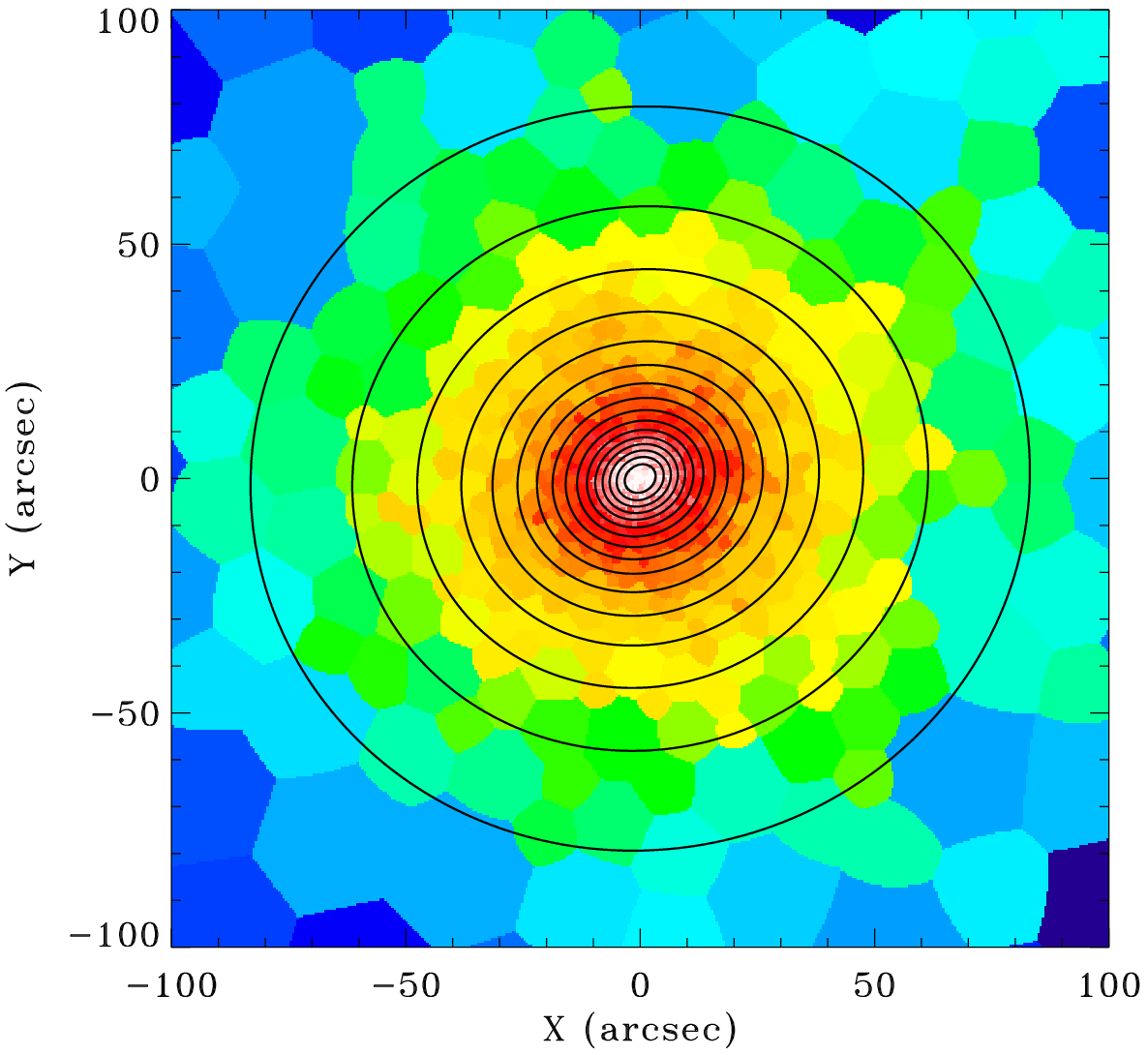}
\caption{{\it Left panel}: Gas X-ray surface brightness (in ${\rm
photons}\,{\rm s}^{-1}\,{\rm cm}^{-2}\,{\rm arcsec}^{-2}$),
ellipticity and position angle profile for NGC~4649. Solid lines
indicate the Chebyshev polynomial fits. {\it Right panel}: Adaptively
binned gas map of NGC~4649 with the fitted elliptical isophotes
overlaid. \label{f.smoothmodel}}
\end{figure*}

\begin{figure*}
\begin{center}
\includegraphics[width=0.3\textwidth]{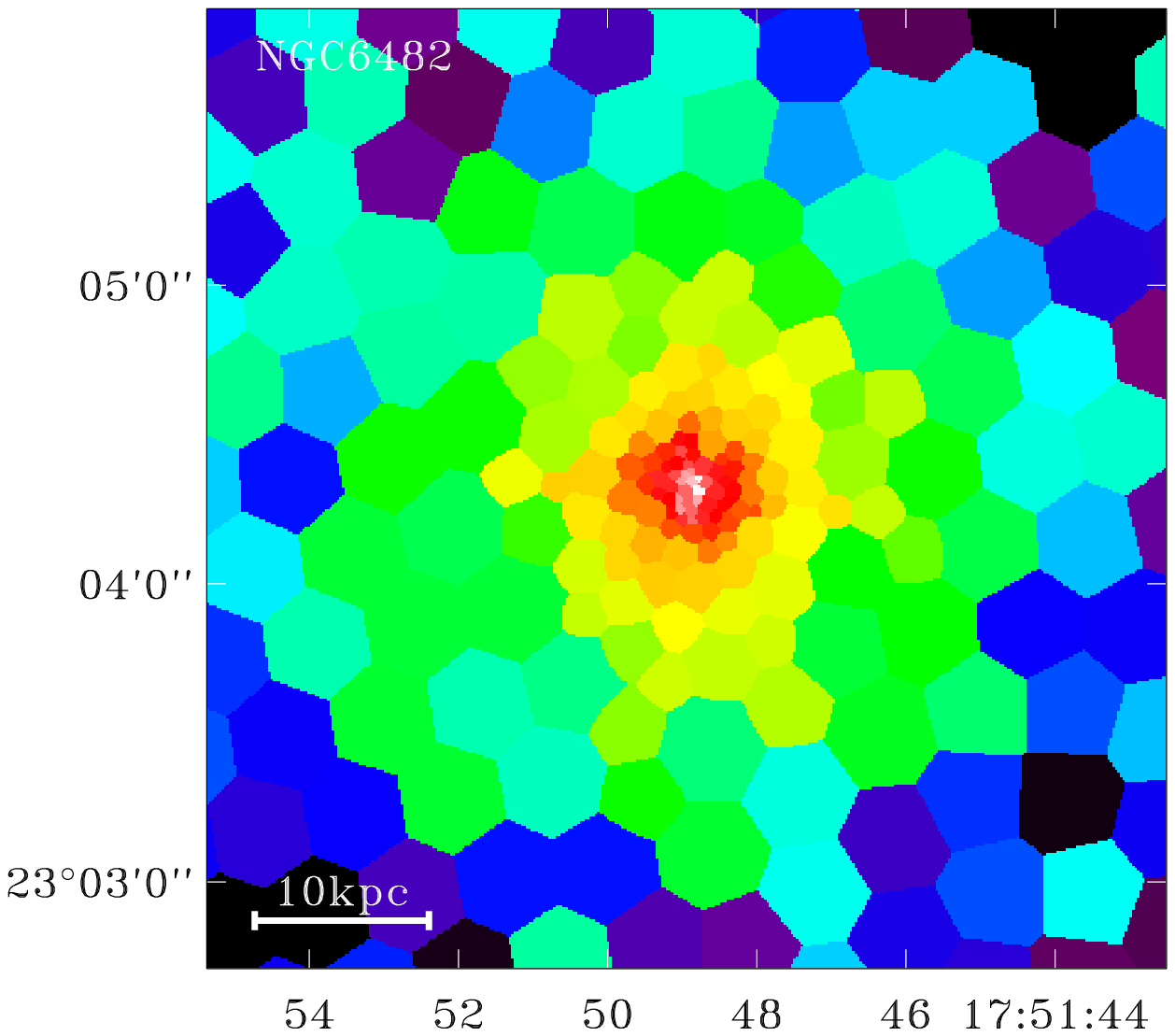}
\includegraphics[width=0.3\textwidth]{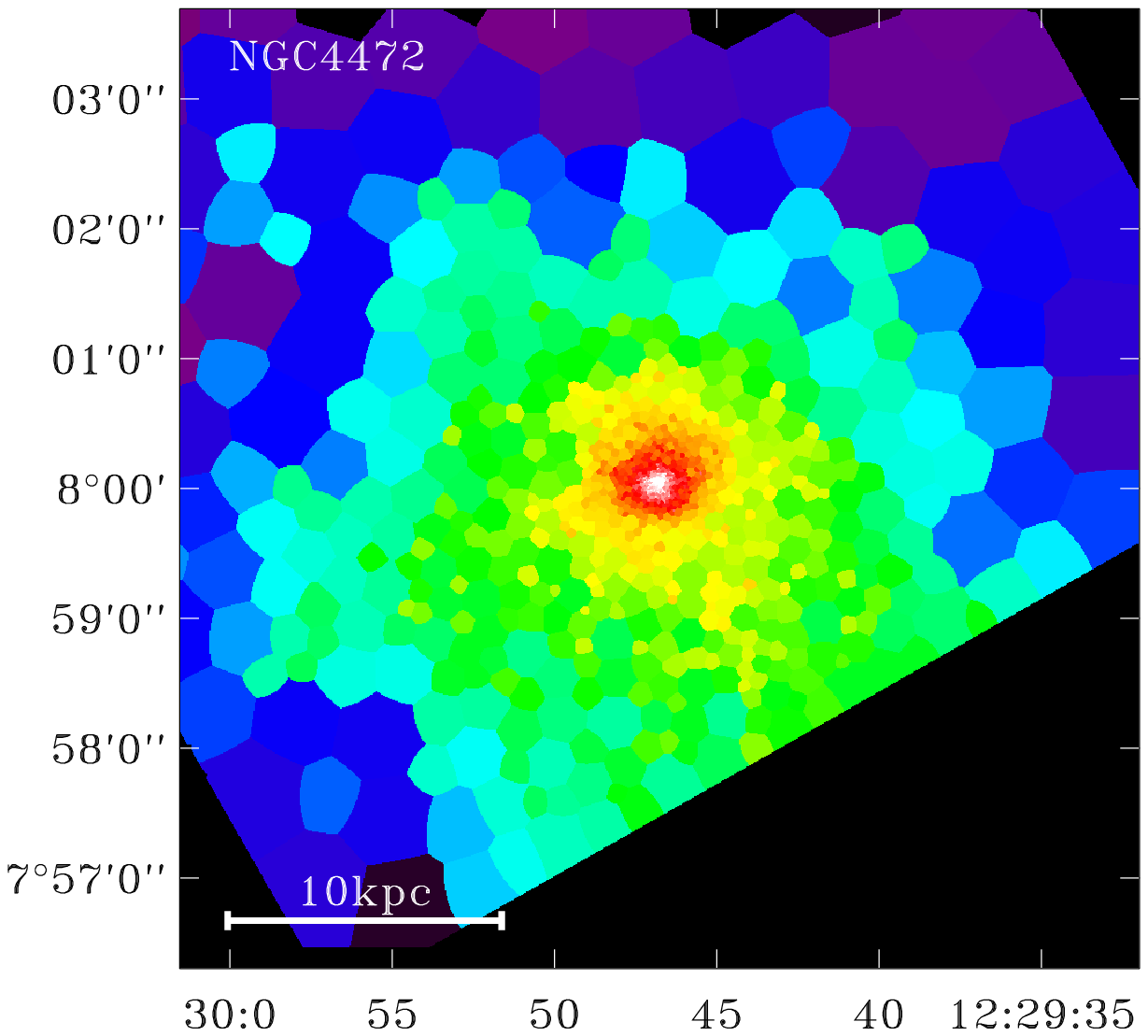}
\includegraphics[width=0.3\textwidth]{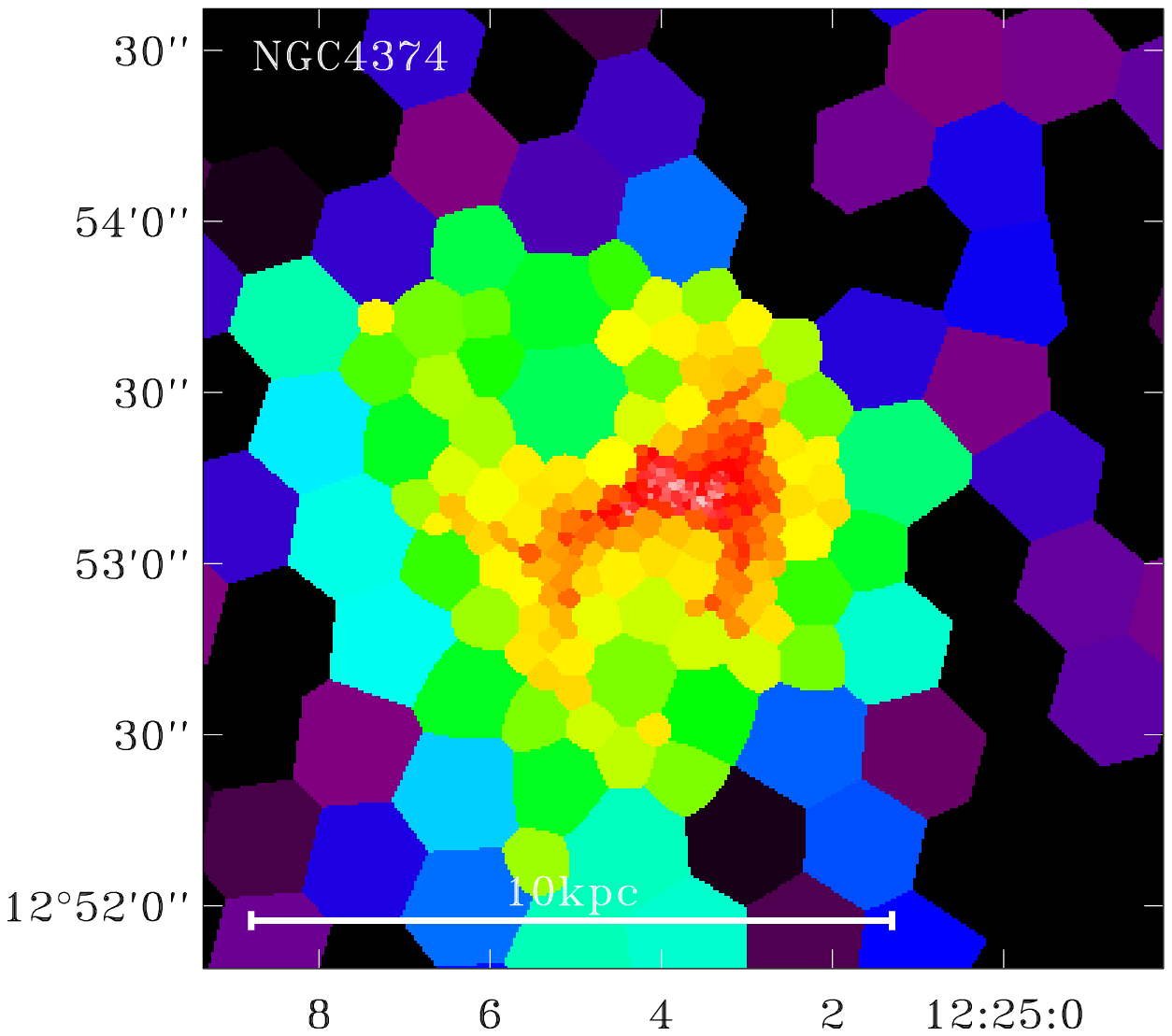}
\end{center}
\caption{Gas surface brightness maps for three elliptical galaxies,
showing the range in asymmetry index $\eta$; Left: NGC~6482
($\eta=0.032$); Center: NGC~4472 ($\eta=0.077$); Right: NGC~4374
($\eta=0.478$). Colors depict gas surface brightness levels with the
color scale ranging from $5\times 10^{-10}$ to $10^{-6} {\rm
photons\,s^{-1}\,cm^{-2}\,arcsec^{-2}}$. The axes are labelled
according to right ascension and declination (2000).\label{f.asymexample}}
\end{figure*}

\subsection{Summary of Paper I and Preliminary Analysis}\label{s4.data}

Our full sample, as described in Paper I, consists of
54 early-type galaxies observed with the the ACIS-S instrument on the
{\it Chandra} satellite during cycles 1-4. The data have been
homogeneously reprocessed to avoid problems due to changes in the
standard calibration pipeline over time. We will briefly summarize the
generation of those data products essential for the understanding of
this paper and refer the interested reader back to Paper I for more
details.

In Paper I, we presented a new method to isolate the hot gas emission
in elliptical galaxies from the contamination of unresolved point
sources. This method is based on the fact that point sources and hot
gas contribute differently to soft ($0.3-1.2\kev$) and hard
($1.2-5\kev$) bands, due to their intrinsically different
spectra. Thus, by subtracting a properly scaled version of the hard
band image from the soft band after removing resolved sources,
we are able to remove the contribution
of the harder unresolved point sources and isolate the soft gas
emission self-consistently, correcting for the amount of gas flux
subtracted from the hard band as well. All images are photon-flux
calibrated, corrected for exposure map effects and background. We
adaptively bin the gas maps with an adaptive binning technique
using weighted Voronoi tesselations
\citep{DiehlWVT,Cappellari}\footnote{http://www.phy.ohiou.edu/$\sim$diehl/WVT}.

For a subset of 36 galaxies with sufficient signal, we
characterize the overall shape of the gas by deriving ellipticity and
position angle profiles. A qualitative comparison with optical DSS-2
$R$-band images and a quantitative comparison with published optical
surface photometry reveal little correlation between optical and X-ray
properties. In particular, we find no correlations between the optical
and X-ray ellipticities at any radius. Even within one effective
radius, where stars dominate the gravitational potential
\citep[e.g.][]{MamonDarkmatterI,HumphreyDarkmatter}, the two are
completely uncorrelated. The second column of Table \ref{t.asymgalaxyprop} lists mean gas ellipticities in the annulus between $0.6$ and $0.9$ $J$-band effective radii, from Paper I. These findings establish that
the gas in elliptical galaxies is, in general, not in precise hydrostatic
equilibrium, at least at a level that precludes using the gas isophotes to constrain the
shapes of dark matter halos
\citep[][]{BuoteGeomtest,BuoteNGC720}. Instead of being
hydrostatically calm, the gas morphology is almost always disturbed,
even for rather X-ray faint galaxies.

In this paper, we restrict our analysis to the subset of 36 galaxies
(Table \ref{t.asymgalaxyprop}) for which we derive ellipticity
profiles in Paper I. Following the technique used in Paper I to
derive photon-flux calibrated images, we also produce calibrated energy-flux images, which are
necessary to determine the central X-ray AGN luminosity
(\S\ref{s4.xrayagn}). We create mono-energetic
exposure maps in steps of 7 in PI ($\sim 100\ev$) and generate counts
images in each individual PI channel ($14.6\ev$-wide). We then divide
each counts image by the energetically closest exposure map to create
a photon-flux-calibrated ``slice,'' and multiply each slice by the
appropriate energy of the PI channel. Finally, we sum all
energy-weighted slices to produce a calibrated energy-flux image that
automatically adjusts the effective exposure map to spatial changes in
the spectral composition of the galaxies, due to radial temperature
gradients and/or point source or AGN contributions. This energy-flux
calibrated image can be used to derive flux values, upper limits, and
flux-calibrated surface brightness profiles even in situations with
insufficient signal to perform spectral fits.

\begin{deluxetable*}{l c rrrrrr}
\tablewidth{0pt}
\tablecaption{Chandra X-ray gas morphology, AGN luminosity and temperature gradient\label{t.asymgalaxyprop}}
\tablehead{
\colhead{Name} & \colhead{} &
\colhead{$\epsilon_{\rm X}$\tablenotemark{a}} & \colhead{$\eta$\tablenotemark{b}} & \colhead{$r_{\rm min}$\tablenotemark{b}} & \colhead{$r_{\rm max}$\tablenotemark{b}} & \colhead{$L_{\rm X,AGN}$\tablenotemark{c}} & \colhead{$\alpha_{24}$\tablenotemark{d}}}
\startdata
IC1262 &  & $ 0.43 \pm 0.09 $ & $ 0.110 \pm 0.042 $ & $   6.4 $ & $ 101.3 $ & $ <  1.9\times 10^{39} $ & $  0.21 \pm  0.07 $\\
IC1459 &  & $ 0.35 \pm 0.12 $ & $ 0.130 \pm 0.040 $ & $   0.7 $ & $   5.1 $ & $  6.4\pm 3.3\times 10^{40} $ & $ -0.27 \pm  0.04 $\\
IC4296 &  & $ 0.28 \pm 0.06 $ & $ 0.133 \pm 0.032 $ & $   1.1 $ & $   5.4 $ & $  3.4\pm 1.1\times 10^{40} $ & $  0.08 \pm  0.07 $\\
NGC0193 &  & $ 0.33 \pm 0.42 $ & $ 0.244 \pm 0.074 $ & $   3.6 $ & $  29.7 $ & $  1.2\pm 0.4\times 10^{40} $ & \nodata\\
NGC0315 &  & $ 0.11 \pm 0.14 $ & $ 0.138 \pm 0.059 $ & $   1.5 $ & $  11.3 $ & $  4.8\pm 4.2\times 10^{41} $ & \nodata\\
NGC0383 &  & $ 0.23 \pm 0.09 $ & $ 0.394 \pm 0.065 $ & $   1.1 $ & $   5.8 $ & $  1.9\pm 0.6\times 10^{40} $ & $  0.50 \pm  0.16 $\\
NGC0404 &  & \nodata & \nodata & \nodata & \nodata & $ <  4.0\times 10^{37} $ & \nodata\\
NGC0507 &  & $ 0.13 \pm 0.17 $ & $ 0.334 \pm 0.055 $ & $   2.2 $ & $  34.7 $ & $ <  6.7\times 10^{38} $ & $  0.16 \pm  0.10 $\\
NGC0533 &  & $ 0.33 \pm 0.04 $ & $ 0.131 \pm 0.032 $ & $   1.3 $ & $  19.7 $ & $  1.0\pm 0.8\times 10^{39} $ & \nodata\\
NGC0720 &  & $ 0.06 \pm 0.07 $ & $ 0.106 \pm 0.042 $ & $   0.7 $ & $  14.3 $ & $ <  2.8\times 10^{38} $ & $ -0.04 \pm  0.15 $\\
NGC0741 &  & $ 0.24 \pm 0.05 $ & $ 0.167 \pm 0.038 $ & $   1.4 $ & $   8.5 $ & $ <  5.6\times 10^{39} $ & $ -0.10 \pm  0.12 $\\
NGC0821 &  & \nodata & \nodata & \nodata & \nodata & $  3.2\pm 1.5\times 10^{38} $ & \nodata\\
NGC1132 &  & $ 0.08 \pm 0.25 $ & $ 0.050 \pm 0.047 $ & $   2.2 $ & $  75.5 $ & $ <  2.3\times 10^{39} $ & $ -0.22 \pm  0.11 $\\
NGC1265 &  & \nodata & \nodata & \nodata & \nodata & $  3.0\pm 1.0\times 10^{40} $ & \nodata\\
NGC1316 &  & $ 0.42 \pm 0.04 $ & $ 0.242 \pm 0.041 $ & $   0.4 $ & $   6.1 $ & $  2.3\pm 1.0\times 10^{38} $ & $  0.12 \pm  0.26 $\\
NGC1399 &  & $ 0.17 \pm 0.07 $ & $ < 0.095 $ & $   0.3 $ & $  12.5 $ & $  1.5\pm 1.3\times 10^{38} $ & $  0.10 \pm  0.03 $\\
NGC1404 &  & $ 0.03 \pm 0.04 $ & $ 0.028 \pm 0.013 $ & $   0.3 $ & $   8.6 $ & $  1.5\pm 0.9\times 10^{38} $ & $  0.15 \pm  0.05 $\\
NGC1407 &  & $ 0.07 \pm 0.03 $ & $ 0.174 \pm 0.023 $ & $   0.7 $ & $   9.8 $ & $  2.0\pm 0.5\times 10^{39} $ & $  0.46 \pm  0.23 $\\
NGC1549 &  & $ 0.24 \pm 0.21 $ & \nodata & \nodata & \nodata & $  4.4\pm 2.0\times 10^{38} $ & \nodata\\
NGC1553 &  & $ 0.57 \pm 0.14 $ & $ 0.174 \pm 0.091 $ & $   1.6 $ & $  11.0 $ & $  5.2\pm 1.1\times 10^{39} $ & $ -0.28 \pm  0.12 $\\
NGC1600 &  & $ 0.24 \pm 0.09 $ & $ 0.114 \pm 0.074 $ & $   1.0 $ & $   9.3 $ & $  1.1\pm 0.8\times 10^{39} $ & \nodata\\
NGC1700 &  & $ 0.31 \pm 0.11 $ & $ 0.154 \pm 0.069 $ & $   1.1 $ & $  24.5 $ & $ <  1.9\times 10^{39} $ & \nodata\\
NGC2434 &  & \nodata & \nodata & \nodata & \nodata & $  1.2\pm 0.7\times 10^{38} $ & \nodata\\
NGC2865 &  & \nodata & \nodata & \nodata & \nodata & $ <  2.6\times 10^{39} $ & \nodata\\
NGC3115 &  & \nodata & \nodata & \nodata & \nodata & $ <  8.5\times 10^{37} $ & \nodata\\
NGC3377 &  & \nodata & \nodata & \nodata & \nodata & $ <  2.2\times 10^{38} $ & \nodata\\
NGC3379 &  & \nodata & \nodata & \nodata & \nodata & $ <  5.0\times 10^{38} $ & \nodata\\
NGC3585 &  & \nodata & \nodata & \nodata & \nodata & $  1.6\pm 1.1\times 10^{38} $ & \nodata\\
NGC3923 &  & $ 0.15 \pm 0.10 $ & $ 0.044 \pm 0.033 $ & $   0.4 $ & $   5.4 $ & $ <  2.5\times 10^{37} $ & $ -0.37 \pm  0.11 $\\
NGC4125 &  & $ 0.22 \pm 0.05 $ & $ 0.056 \pm 0.027 $ & $   0.7 $ & $  13.0 $ & $  2.2\pm 0.9\times 10^{38} $ & $ -0.15 \pm  0.16 $\\
NGC4261 &  & $ 0.08 \pm 0.08 $ & $ 0.142 \pm 0.054 $ & $   0.6 $ & $   4.3 $ & $  1.1\pm 0.2\times 10^{40} $ & \nodata\\
NGC4365 &  & \nodata & \nodata & \nodata & \nodata & $ <  1.4\times 10^{38} $ & \nodata\\
NGC4374 &  & $ 0.31 \pm 0.04 $ & $ 0.478 \pm 0.050 $ & $   0.4 $ & $   5.0 $ & $  8.2\pm 2.6\times 10^{38} $ & $  0.44 \pm  0.08 $\\
NGC4406 &  & $ 0.21 \pm 0.09 $ & $ 0.130 \pm 0.021 $ & $   0.9 $ & $   9.2 $ & $  8.6\pm 1.4\times 10^{38} $ & $ -0.05 \pm  0.04 $\\
NGC4472 &  & $ 0.05 \pm 0.03 $ & $ 0.077 \pm 0.012 $ & $   0.3 $ & $  10.0 $ & $  9.6\pm 4.1\times 10^{37} $ & \nodata\\
NGC4494 &  & \nodata & \nodata & \nodata & \nodata & $ <  1.9\times 10^{39} $ & \nodata\\
NGC4526 &  & $ 0.07 \pm 0.16 $ & $ 0.258 \pm 0.177 $ & $   0.5 $ & $   2.9 $ & $  7.9\pm 4.6\times 10^{38} $ & \nodata\\
NGC4552 &  & $ 0.31 \pm 0.02 $ & $ 0.149 \pm 0.028 $ & $   0.3 $ & $   3.9 $ & $ <  7.2\times 10^{38} $ & $  0.42 \pm  0.16 $\\
NGC4555 &  & \nodata & \nodata & \nodata & \nodata & $ <  4.2\times 10^{40} $ & $ -0.02 \pm  0.03 $\\
NGC4564 &  & \nodata & \nodata & \nodata & \nodata & $ <  4.4\times 10^{38} $ & \nodata\\
NGC4621 &  & \nodata & \nodata & \nodata & \nodata & $  6.2\pm 3.6\times 10^{38} $ & \nodata\\
NGC4636 &  & $ 0.22 \pm 0.02 $ & $ 0.086 \pm 0.009 $ & $   0.3 $ & $   9.5 $ & $  6.2\pm 1.4\times 10^{37} $ & $  0.04 \pm  0.02 $\\
NGC4649 &  & $ 0.04 \pm 0.03 $ & $ 0.049 \pm 0.035 $ & $   0.3 $ & $   7.7 $ & $  3.0\pm 2.7\times 10^{37} $ & $ -0.01 \pm  0.01 $\\
NGC4697 &  & $ 0.20 \pm 0.16 $ & $ < 0.080 $ & $   0.7 $ & $   4.2 $ & $ <  1.6\times 10^{39} $ & \nodata\\
NGC5018 &  & $ 0.23 \pm 0.26 $ & \nodata & \nodata & \nodata & $ <  3.3\times 10^{38} $ & \nodata\\
NGC5044 &  & $ 0.46 \pm 0.05 $ & $ 0.213 \pm 0.022 $ & $   1.1 $ & $  19.8 $ & $  8.0\pm 2.6\times 10^{38} $ & $  0.12 \pm  0.02 $\\
NGC5102 &  & \nodata & \nodata & \nodata & \nodata & $ <  3.1\times 10^{36} $ & \nodata\\
NGC5171 &  & \nodata & \nodata & \nodata & \nodata & $ <  6.3\times 10^{38} $ & \nodata\\
NGC5532 &  & \nodata & \nodata & \nodata & \nodata & $  6.9\pm 3.5\times 10^{40} $ & $ -0.38 \pm  0.07 $\\
NGC5845 &  & \nodata & \nodata & \nodata & \nodata & $  6.6\pm 3.3\times 10^{38} $ & \nodata\\
NGC5846 &  & $ 0.31 \pm 0.02 $ & $ 0.141 \pm 0.018 $ & $   0.7 $ & $  13.2 $ & $  2.8\pm 0.8\times 10^{38} $ & $  0.03 \pm  0.17 $\\
NGC6482 &  & $ 0.14 \pm 0.05 $ & $ < 0.032 $ & $   1.2 $ & $  30.3 $ & $ <  2.5\times 10^{39} $ & $ -0.34 \pm  0.02 $\\
NGC7052 &  & $ 0.26 \pm 0.12 $ & $ 0.118 \pm 0.067 $ & $   1.7 $ & $   5.5 $ & $  1.4\pm 0.5\times 10^{40} $ & \nodata\\
NGC7618 &  & $ 0.38 \pm 0.14 $ & $ 0.225 \pm 0.060 $ & $   3.4 $ & $  46.8 $ & $  3.1\pm 1.7\times 10^{39} $ & $ -0.12 \pm  0.08 $\\
\enddata
\tablenotetext{a}{Mean X-ray gas ellipticity $\epsilon_{\rm X}$ between $0.6-0.9\, R_{\rm J}$ \citep[$J$-band effective radius from 2MASS extended source catalog,][]{2MASS}.}
\tablenotetext{b}{Asymmetry index $\eta$, measured between $r_{\min}$ and $r_{\rm max}$ (in $\kpc$).}
\tablenotetext{c}{X-ray AGN luminosity (in ${\rm ergs\,s^{-1}}$) between $0.3-5\kev$, derived from $\beta$ and S\'ersic model plus point source fits to radial surface brightness profiles.}
\tablenotetext{d}{Mean logarithmic temperature gradient $d\ln T/d \ln r $ evaluated at radii between 2 and 4 $R_J$}
\end{deluxetable*}

\subsection{Quantifying Asymmetry}\label{s4.asymmetry}

\subsubsection{Definition of the Asymmetry Index $\eta$}

Since many of the {\it Chandra} data sets lack the required signal and
resolution to reliably identify individual asymmetric features in the
gas surface brightness maps, we are forced to measure asymmetry
statistically. We first construct a smooth, symmetric surface
brightness model to subtract from the gas map. We base this model on
the isophotal profiles computed in Paper I. These profiles give
surface brightness ($I_{\rm X}$), ellipticity ($\epsilon_{\rm X}$), and
major axis position angle ($PA_{\rm X}$) as a function of mean radius
$r$. We fit this surface brightness profile in $\log r - \log I_{\rm
X}$ space with a quadratic Chebyshev polynomial. For two galaxies with
more complex surface brightness profiles (NGC~1316 and NGC~4636), we use a third order
polynomial, and for one other object (NGC~4472), we resort to fourth order. To
accommodate moderate changes in ellipticity and isophotal twists, we
fit the ellipticity and position angle profiles in $\log r -
\epsilon_{\rm X}$ and $\log r - {\rm PA}_{\rm X}$ space with
straight lines. From the fits, we compute a smooth, symmetric model for
the gas distribution. The left panel of Figure \ref{f.smoothmodel}
shows all three profile fits for NGC~4649. The adaptively binned gas
map in the right panel has the elliptical isophotes of the adopted
fits overlaid. Finally, we bin the surface brightness model to match
the binning structure of the adaptively binned gas map and subtract
the binned smooth model to reveal small scale asymmetries in the gas
distribution.

To quantify the degree of asymmetry in this residual map, we define
the asymmetry index $\eta$ in the following way:
\begin{equation}\label{e.asymindex}
\eta={{1}\over{N}}\, \sum_{i=1}^{N}{\left[{\left({G_i-M_i\over
M_i}\right)^2 - \left({\sigma_{G,i}\over M_i}\right)^2}\right]}.
\end{equation}
The asymmetry index is the sum over all pixels $i$ of the squared
relative deviations of the binned gas image $G_i$ from the smooth
model $M_i$, over and above the expected statistical deviations due to
Poisson noise $\sigma_{G,i}$\footnote{Equation (\ref{e.asymindex}) is
equivalent to an area-weighted average over bins. We find empirically
that an unweighted average gives too much emphasis to bright regions
and small-scale structure.}. We find that $\eta$ is an unbiased
measure of asymmetry, independent of exposure time, radial fitting
range, background level or signal-to-noise ratio. The errors on the
$\eta$ values are estimated from a bootstrap analysis of 20
Monte-Carlo simulations.

Figure \ref{f.asymexample} shows representative examples spanning the
range of $\eta$ values present in our sample. NGC~6482 ($\eta < 0.032$,
left) shows no apparent signs of disturbance and is statistically
consistent with a smooth, purely elliptical model
\citep{PonmanNGC6482}. In NGC~4472 ($\eta=0.077$, center), one can
detect low-level surface brightness depressions at several radii
\citep{BillerNGC4472}. NGC~4374 \citep[right;][]{Fin01} has the
highest $\eta$ value ($\eta=0.48$) in our sample. This sequence can
serve as a guide for the meaning of $\eta$. However, one should keep in
mind that the examples in Figure \ref{f.asymexample} have been chosen
to have high signal-to-noise ratios for display purposes. Most galaxies
have significantly less signal and the asymmetric structures cannot be
picked out by eye as easily.

The third, fourth, and fifth columns of Table \ref{t.asymgalaxyprop} list the values of $\eta$ and the range of radii from which they are derived. We have chosen to calculate $\eta$ from the full radial range over which isophote fits are possible for each object, in order to take advantage of as much detected signal as possible. This choice unavoidably creates a certain inhomogeneity in the sample since $\eta$ is not always being measured at the same physical scale. However, we have tested the sensitivity of all of the results of this paper to the size of the extraction annulus by separating into subsamples with small and large outer radii $r_{\rm max}$. In no case do we find any statistically significant difference between the subsamples.

\subsubsection{Simulations and Tests}\label{s4.asymmetrytest}

\begin{figure}
\begin{center}
\includegraphics[width=0.45\textwidth]{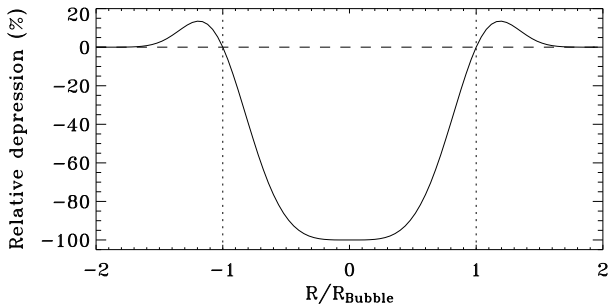}
\end{center}
\caption{One-dimensional cut through a
simulated ``bubble depression'', demonstrating the fractional
depression strength of the bubble with respect to its radial
extent. The bubble has negative surface brightness deviations within
the bubble radius $R_{\it Bubble}$, and slightly enhanced rims around
it. The profile shape is chosen such that the integrated deviations
integrate to 0 over area. \label{f.bubbleprofile}}
\end{figure}

To test the behavior of the asymmetry index $\eta$, we simulate images
that qualitatively reproduce the variety of gas morphologies seen in
the data. We start with a $256\times 256$ pixel realization of a
S\'ersic model with index $n=4$ and a half-light radius of 50
pixels. We then add disturbances as relative surface brightness
depressions surrounded by enhanced rim emission, mimicking the
appearance of AGN-induced ``bubbles'', often observed in clusters of
galaxies \citep[e.g.][]{McnamaraHydraA}. Figure \ref{f.bubbleprofile}
shows a one-dimensional cut through our adopted circular bubble
template profile, which is obtained by subtracting two
$e^{-(r/\sigma)^4}$ functions with slightly different $\sigma$ values,
and adjusting their normalizations such that integrating the 2-d
profile yields a total deviation of zero. We define the ``depression
strength'' as the maximum relative deviation at the bubble center,
indicated by the $-100\%$ mark in our template, while the dashed
vertical lines in Figure \ref{f.bubbleprofile} mark the ``bubble
size''. We then control the level of asymmetry in our simulations by
varying the depression strength, bubble size, and the total number of
bubbles.

We simulate the disturbed gas images by randomly distributing the
bubbles over the entire S\'ersic model image. Then, following our real
data analysis, we construct a smooth model to subtract. For
simplicity, we extract a radial surface brightness profile from
circular annuli, and fit it with a S\'ersic model, instead of
constructing a fully general elliptical model. We adaptively bin the
image to a signal-to-noise per bin of 4, and apply the same binning
structure to the smooth model. After subtracting the binned model, we
derive the asymmetry index from the residuals according to equation
(\ref{e.asymindex}). We produce a suite of test simulations, keeping
two parameters fixed while varying the third. For small numbers of
simulated bubbles, the disturbed surface brightness models resemble
galaxies in which one is able to reliably identify the location of
individual bubbles. Although very large numbers of simulated bubbles
are unrealistic for real galaxies, we increase the number of bubbles
beyond the physically motivated level to achieve more complex
asymmetries in the morphology.

\begin{figure}
\begin{center}
\includegraphics[width=0.45\textwidth]{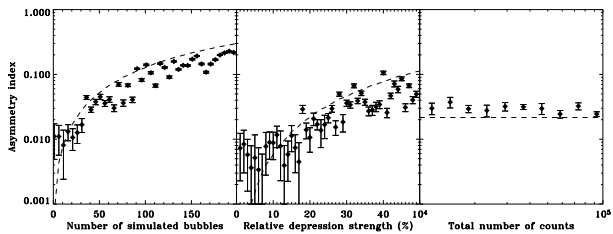}
\end{center}
\caption{Tests of the asymmetry index using models with
simulated bubble-like depressions (see
Fig. \ref{f.bubbleprofile}). The asymmetry index $\eta$ is sensitive
to the number of simulated bubbles (left), and the relative strength of
the depression (middle), but not to the effective exposure time, characterized
by the total number of counts (right). The dashed lines mark the
expected behavior for perfect (noise-free) data. \label{f.asymtest}}
\end{figure}

We find that $\eta$ is largely independent of the bubble size, but
correlates positively with the number of simulated bubbles and their
relative depression strength, as shown in the left and middle panel of
Figure \ref{f.asymtest}. We compute the expected trends by simulating
20 random spatial configurations for each parameter set without adding
Poisson noise, and computing the average of all $\eta$ values of these
``perfect'' data sets. We find that our asymmetry index values are
fair representations of the expected values.

The asymmetry index is constructed to account for the statistical
scatter in the data and to be insensitive to differences in data
quality. To verify this, we repeat our tests, keeping the bubble
parameters fixed while varying the total galaxy luminosity, background
level, cutoff radius, and the exposure time of the observation. We
find no correlations with any of these parameters. As an example, in
the right panel of Figure \ref{f.asymtest} we show one test in which
we vary the exposure time. One can see that $\eta$ is not affected by
the increasing resolution that is a result of the rising total number
of counts. We conclude that $\eta$ is an unbiased measure of the
asymmetry present in the gas maps, and unaffected by non-morphological
data properties.

\subsection{X-ray AGN Luminosities\label{s4.xrayagn}}

We identify the location of the central AGN in the X-ray image by
overlaying the high-accuracy center from the 2MASS extended source
catalogue \citep{2MASS}. The 2MASS position is derived from the
luminosity weighted center of the co-added $K$, $J$ and $H$ band
images and has a typical $1\sigma$-uncertainty of only
$0.3\arcsec$. We combine this uncertainty with {\it Chandra}'s
pointing
accuracy\footnote{http://cxc.harvard.edu/mta/ASPECT/abs\_point.html},
and manually assign the AGN to the brightest X-ray point source in the
$0.3-5.0\kev$ counts image within a $\sim 3$ pixel radius of the 2MASS
center. If multiple source candidates are within this field, we take
the point source closest to the centroid of the diffuse emission. In cases
where soft gas emission completely dominates the central region, we
revert to the $2-5\kev$ band for identification instead. This
technique allows a rather reliable AGN identification, with only few
ambiguous cases.

Unfortunately, the location of the AGN coincides with the peaks of
both the diffuse gas emission and the stellar point source
component. While this sometimes helps in pinpointing the location of the AGN, it
makes it difficult to determine the AGN flux. The standard technique
is to extract a spectrum from the central region and to attempt to
spectrally disentangle the AGN component from the contaminating
diffuse gas and low-mass X-ray binaries (LMXBs) \citep[e.g.][]{KimNGC1316}. However, the weak
AGN in normal elliptical galaxies often yield insufficient counts to
constrain all parameters in this complex spectral fit. This fit is
additionally complicated by the fact that the AGN and unresolved LMXB
components are both well-described by featureless power-laws, and can
sometimes become virtually indistinguishable. For many observations in
our sample, the AGN contributes no more than a few counts, making a
reliable spectral fit practically impossible.

Instead, we decide to spatially disentangle the X-ray AGN emission
from its complex background using the energy-flux calibrated images of
diffuse emission. The energy weighting makes it easier to distinguish
the hard AGN component. We bin the energy-flux image into circular
annuli centered on the AGN position and adaptively change the annular
bin sizes to enforce a minimum signal-to-noise requirement of 2. As
this would result in very large numbers of bins at larger radii, we
also require the annular width to be larger than 10\% of the mean bin
radius to ensure proper azimuthal averaging within a bin, generally resulting in
a logarithmic binning at large radii.

\begin{figure*}
\begin{center}\includegraphics[width=0.75\textwidth]{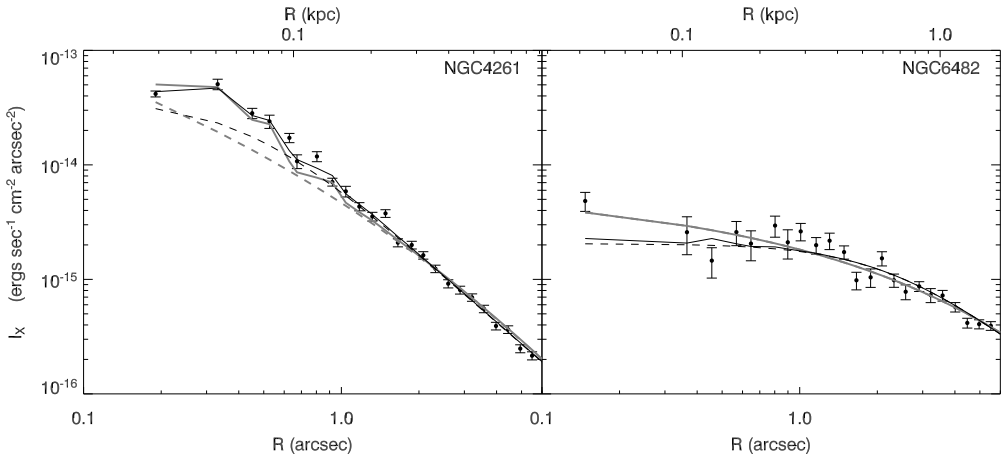}
\end{center}
\caption{Radial X-ray surface brightness profiles for the inner
$6\arcsec$ of NGC~4261 (left) and NGC~6482 (right). Thick solid lines
indicate S\'ersic model fits, thin solid lines show $\beta$ model
fits, including a PSF model for the central AGN. Dashed lines show the
same model fits without the AGN component. NGC~4261's AGN is detected
at a $7\sigma$ level, NGC~6482 profile reveals no AGN signature. For
NGC~6482, the AGN flux yielded by the S\'ersic model fit is so small
that the thick dashed line is invisible, as it is covered by the thick
solid line.\label{f.agnfit}}
\end{figure*}

We fit the inner 32 arcsec of the radial surface brightness profile
with a two-component model, one representing the point source and one
the diffuse emission. The point source profile is obtained from a
normalized mono-energetic point-spread function (PSF) centered on the
position of the central AGN and computed at its mean photon energy. We
bin the PSF image to the same circular binning, to give a radial point source
model that matches our binning structure with only the normalization flux as a free parameter. To
represent the diffuse emission we use a S\'ersic or a $\beta$ model,
on top of a uniform background. Figure \ref{f.agnfit} shows the inner
10~arcsec of two galaxies with a statistically significant AGN
detection (NGC~4261, left) and a non-detection (NGC~6482, right). The
dashed lines indicate the diffuse component from the S\'ersic (grey)
and $\beta$ model fit (black), while the solid lines show the same
fits with the additional AGN component. All fits are inspected by eye
and obviously unconverged fits are repeated over a different radial
extent until a satisfactory fit is obtained.

Even though both models generally produce statistically equally good fits, the
$\beta$ model fits systematically yield higher fluxes for the AGN, as
the $\beta$-profile is flatter at small radii. We adopt the average of
the AGN fluxes yielded by the $\beta$ and S\'ersic model fits as the
best value. The difference between the two individual fits and the
average value is assumed to represent a $3\sigma$ systematic error and
is combined with the statistical errors yielded by each profile
fitting procedure. Finally, we correct the AGN fluxes for Galactic
absorption with the same correction factor that is used to correct the
total gas flux, computed from integrating the best spectral fit for
the gas emission with and without the Galactic absorption factor. This
correction may be a slight overestimate, since the AGN emission is
generally harder and less strongly affected by absorption; on the
other hand the column density is also likely to be higher due to
intrinsic absorption inside the galaxy. This correction is generally
rather small, well below our statistical and systematical errors, and
does not affect the computed X-ray AGN luminosity significantly.

The final AGN luminosities $L_{\rm X,AGN}$ are reported in the sixth column of Table
\ref{t.asymgalaxyprop}, together with the combined statistical and
systematic errors. Flux values that are detected below the $1\sigma$
confidence limit are reported as upper limits. For objects with
insufficient signal to produce a surface brightness profile, we sum up
the energy-flux calibrated image within the region containing 99\% of
the PSF flux and correct for a flat background extracted from an
adjacent annulus extending out to 10~arcsec. Since we have no way to
distinguish between a central peak in gas emission, LMXBs, or an
actual AGN for these galaxies, we quote the $3\sigma$ upper bound of
the summed flux as an upper limit on their AGN luminosity.

\subsection{Radio Luminosities}

We use the NRAO VLA Sky Survey \citep[NVSS;][]{NVSS} to derive homogeneous $20\cm$ radio continuum luminosities for our sample.  As we are interested in the AGN's impact on the ISM, we sum up all associated radio sources within 3 J-band effective radii from the 2MASS catalog \citep{2MASS} radii to derive the NVSS radio luminosity $L_{\rm NVSS}$, thus including contributions from extended structures. This results in general in an extraction radius of approximately $1.5\arcmin$ for most of the sample (see Paper I for a complete list of optical radii). The  formal average NVSS source density for the full NVSS catalog is about 50 sources per square degree \citep{NVSS}, resulting in a probability to include unrelated background sources within $1.5\arcmin$ of only $9\%$. For 63\% of our sample, this probability lies below 10\%, for 91\% it lies below 20\%. To further reduce the chance of including background sources, we inspect all NVSS images by eye and compare radio features with X-ray and optical images to find possible counterparts. We manually remove obviously non-associated background sources from the source list.

The resulting luminosities  are listed in the second column of Table \ref{t.radiogalaxyprop}. For the majority of our sample, the limited spatial resolution of the NVSS survey precludes reliably distinguishing between extended and point-like radio sources. Only where unambiguously possible, we determine the NVSS position angle by manually measuring the orientation of the major axis.

Twenty-three galaxies in our sample have been observed in the Faint
Images of the Radio Sky at Twenty-Centimeters (FIRST) VLA survey
\citep{FIRST}, which has a much higher spatial resolution than NVSS,
with a similar detection limit. We extract FIRST radio luminosities
from the inner 3 optical effective radii to match the NVSS extraction
region. A comparison between the two radio surveys indicates that both
are yielding consistent values. However, the higher spatial resolution
of the FIRST survey gives us the opportunity to extract fluxes from
the central point source alone, largely excluding contributions from
the radio jets. Thus, instead of using a large extraction radius, we
sum the flux within a smaller 30~arcsec radius to determine the FIRST
luminosity $L_{\rm FIRST}$, listed in the third column of Table
\ref{t.radiogalaxyprop}. These values are systematically smaller than
the NVSS fluxes due to the smaller extraction region, but still very
well correlated. This indicates that our larger NVSS sample is not
heavily contaminated by background objects, but that the extra flux is
most likely associated with extended radio structure intrinsic to the galaxy.

\begin{deluxetable}{l c rrr}
\tablewidth{0pt}
\tablecaption{Radio luminosities and galaxy environment\label{t.radiogalaxyprop}}
\tablehead{
\colhead{Name} & \colhead{\phm} & \colhead{$L_{\rm NVSS}$\tablenotemark{a}} &\colhead{$L_{\rm FIRST}$\tablenotemark{a}} &
\colhead{$\log_{10} \rhomass$\tablenotemark{b}}}
\startdata
IC1262 &  & $  1.8\pm 0.5\times 10^{33} $ & $  2.4\pm 0.8\times 10^{32} $ & $ 3.47 \pm 0.22 $\\
IC1459 &  & $  1.3\pm 0.3\times 10^{33} $ & \nodata & $ 2.41 \pm 0.31 $\\
IC4296 &  & $  6.0\pm 1.8\times 10^{33} $ & \nodata & $ 2.77 \pm 0.25 $\\
NGC0193 &  & $  5.6\pm 1.7\times 10^{33} $ & \nodata & $ 2.66 \pm 0.31 $\\
NGC0315 &  & $  9.8\pm 2.9\times 10^{33} $ & \nodata & $ 2.43 \pm 0.43 $\\
NGC0383 &  & $  1.3\pm 0.4\times 10^{34} $ & \nodata & $ 3.52 \pm 0.13 $\\
NGC0404 &  & $  4.3\pm 0.6\times 10^{28} $ & \nodata & \nodata\\
NGC0507 &  & $  6.1\pm 1.8\times 10^{32} $ & \nodata & $ 3.03 \pm 0.22 $\\
NGC0533 &  & $  2.1\pm 0.6\times 10^{32} $ & \nodata & $ 3.17 \pm 0.19 $\\
NGC0720 &  & $ <  2.3\times 10^{30} $ & \nodata & $ 2.57 \pm 0.25 $\\
NGC0741 &  & $  6.4\pm 1.9\times 10^{33} $ & \nodata & $ 2.95 \pm 0.25 $\\
NGC0821 &  & $ <  1.7\times 10^{30} $ & \nodata & \nodata\\
NGC1132 &  & $  1.1\pm 0.3\times 10^{32} $ & $  3.8\pm 1.2\times 10^{31} $ & $ 3.07 \pm 0.25 $\\
NGC1265 &  & $  3.3\pm 1.0\times 10^{34} $ & \nodata & \nodata\\
NGC1316 &  & $  1.6\pm 0.3\times 10^{32} $ & \nodata & $ 3.13 \pm 0.13 $\\
NGC1399 &  & $  3.0\pm 0.4\times 10^{32} $ & \nodata & $ 3.28 \pm 0.12 $\\
NGC1404 &  & $  2.1\pm 0.5\times 10^{30} $ & \nodata & $ 3.27 \pm 0.11 $\\
NGC1407 &  & $  9.7\pm 2.3\times 10^{31} $ & \nodata & $ 3.10 \pm 0.14 $\\
NGC1549 &  & \nodata & \nodata & $ 2.74 \pm 0.22 $\\
NGC1553 &  & \nodata & \nodata & $ 2.96 \pm 0.18 $\\
NGC1600 &  & $  3.6\pm 1.1\times 10^{32} $ & \nodata & $ 3.09 \pm 0.19 $\\
NGC1700 &  & $ <  8.8\times 10^{30} $ & \nodata & $ 2.31 \pm 0.43 $\\
NGC2434 &  & \nodata & \nodata & $ 2.56 \pm 0.25 $\\
NGC2865 &  & $ <  4.3\times 10^{30} $ & \nodata & \nodata\\
NGC3115 &  & $ <  2.8\times 10^{29} $ & $ <  2.2\times 10^{29} $ & \nodata\\
NGC3377 &  & $ <  3.8\times 10^{29} $ & $ <  2.8\times 10^{29} $ & \nodata\\
NGC3379 &  & $  3.2\pm 0.7\times 10^{29} $ & $ <  2.5\times 10^{29} $ & $ 3.82 \pm 0.10 $\\
NGC3585 &  & $ <  1.2\times 10^{30} $ & \nodata & $ 2.43 \pm 0.31 $\\
NGC3923 &  & $ <  1.6\times 10^{30} $ & \nodata & $ 2.89 \pm 0.16 $\\
NGC4125 &  & $  1.7\pm 0.4\times 10^{31} $ & \nodata & $ 3.13 \pm 0.13 $\\
NGC4261 &  & $  1.0\pm 0.2\times 10^{34} $ & $  1.7\pm 0.3\times 10^{32} $ & $ 3.09 \pm 0.14 $\\
NGC4365 &  & $ <  1.2\times 10^{30} $ & $ <  9.7\times 10^{29} $ & $ 3.16 \pm 0.13 $\\
NGC4374 &  & $  2.5\pm 0.3\times 10^{33} $ & $  3.4\pm 0.5\times 10^{32} $ & $ 3.18 \pm 0.14 $\\
NGC4406 &  & $  1.1\pm 0.2\times 10^{30} $ & $ <  2.6\times 10^{30} $ & $ 3.31 \pm 0.13 $\\
NGC4472 &  & $  8.1\pm 0.8\times 10^{31} $ & $  3.9\pm 0.5\times 10^{31} $ & $ 3.41 \pm 0.12 $\\
NGC4494 &  & $ <  8.7\times 10^{29} $ & $ <  6.4\times 10^{29} $ & \nodata\\
NGC4526 &  & $  6.6\pm 1.2\times 10^{30} $ & $  4.1\pm 0.9\times 10^{30} $ & $ 2.54 \pm 0.31 $\\
NGC4552 &  & $  2.9\pm 0.4\times 10^{31} $ & $  3.2\pm 0.5\times 10^{31} $ & $ 3.62 \pm 0.09 $\\
NGC4555 &  & $ <  2.8\times 10^{31} $ & $  2.2\pm 0.7\times 10^{31} $ & $ 3.19 \pm 0.22 $\\
NGC4564 &  & $ <  6.7\times 10^{29} $ & $ <  5.4\times 10^{29} $ & \nodata\\
NGC4621 &  & $ < 10.0\times 10^{29} $ & $ <  7.6\times 10^{29} $ & $ 2.66 \pm 0.25 $\\
NGC4636 &  & $  2.7\pm 0.3\times 10^{31} $ & $  1.5\pm 0.2\times 10^{31} $ & $ 3.47 \pm 0.12 $\\
NGC4649 &  & $  9.8\pm 1.4\times 10^{30} $ & $  6.4\pm 1.1\times 10^{30} $ & $ 3.39 \pm 0.12 $\\
NGC4697 &  & $ <  4.1\times 10^{29} $ & $ <  3.4\times 10^{29} $ & $ 3.65 \pm 0.11 $\\
NGC5018 &  & $ <  4.8\times 10^{30} $ & \nodata & $ 2.50 \pm 0.31 $\\
NGC5044 &  & $  4.0\pm 1.0\times 10^{31} $ & \nodata & $ 3.17 \pm 0.13 $\\
NGC5102 &  & $  6.1\pm 1.5\times 10^{28} $ & \nodata & \nodata\\
NGC5171 &  & $ <  2.9\times 10^{31} $ & $  2.7\pm 0.8\times 10^{31} $ & \nodata\\
NGC5532 &  & $  5.8\pm 1.7\times 10^{34} $ & $  5.1\pm 1.6\times 10^{33} $ & $ 3.11 \pm 0.25 $\\
NGC5845 &  & $ <  2.0\times 10^{30} $ & $ <  1.5\times 10^{30} $ & \nodata\\
NGC5846 &  & $  1.6\pm 0.3\times 10^{31} $ & $  1.0\pm 0.2\times 10^{31} $ & $ 2.97 \pm 0.15 $\\
NGC6482 &  & $ <  1.0\times 10^{31} $ & \nodata & $ 2.64 \pm 0.31 $\\
NGC7052 &  & $  1.2\pm 0.4\times 10^{33} $ & \nodata & $ 2.72 \pm 0.31 $\\
NGC7618 &  & $  2.7\pm 0.8\times 10^{32} $ & \nodata & $ 2.76 \pm 0.31 $\\
\enddata
\tablenotetext{a}{20 cm continuum radio luminosity (in ${\rm ergs\, s^{-1} \, Hz^{-1}}$) from NVSS within $3\, R_{\rm J}$, and FIRST within $30\arcsec$.}
\tablenotetext{b}{Projected local galaxy density (in $\rm Mpc^{-2}$), derived from the 2MASS extended source catalog \citep{2MASS}. Reported errors only include statistical errors, not systematic errors.}
\end{deluxetable}

\subsection{Correlation between Radio and X-ray AGN Luminosities}\label{s4.agnradioxray}

\begin{figure*}
\begin{center}
\includegraphics[width=0.9\textwidth]{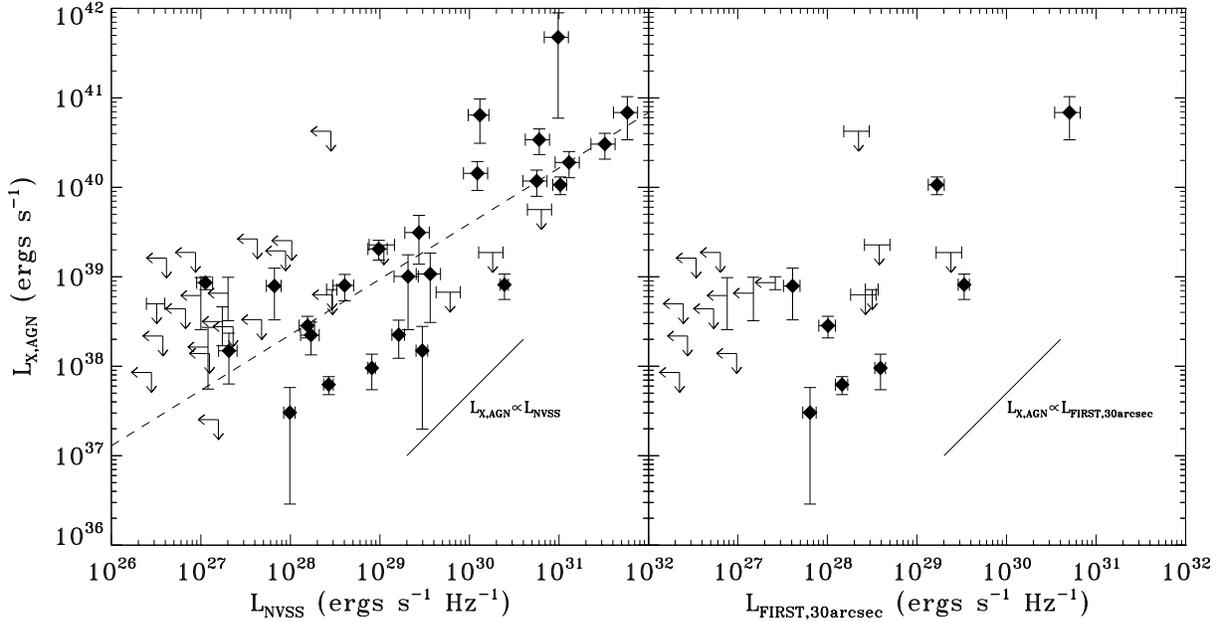}
\end{center}
\caption{X-ray AGN luminosity vs. $20\cm$ continuum radio luminosity with data taken from NVSS within 3 optical radii ({\it left}) and from FIRST ({\it right}) within a $30\arcsec$ radius. Arrows indicate $3\sigma$ upper limits. The slope of the relationship between radio and X-ray luminosity is close to linear, as indicated by the solid lines. The dashed line in the left panel shows the \texttt{bandfit} best fit relation, which has a slope of $0.62\pm0.10$. Thus, X-ray and radio luminosities are each measures of AGN activity.
\label{f.radioxrayagn}}
\end{figure*}

Figure \ref{f.radioxrayagn} compares the AGN X-ray luminosity with both the NVSS (left) and FIRST (right) 20 cm continuum radio powers. Both plots show a clear correlation between radio and X-ray luminosities, spanning almost six orders of magnitude. The slope is consistent with a purely linear relation between both luminosities, as indicated by the solid lines. We analyze all of the correlations in this paper using an algorithm (``\texttt{bandfit}''; see Appendix \ref{s.bandfit}) that models the distribution of data points as a linear band with finite Gaussian intrinsic width. \texttt{Bandfit} can handle data with errors in either or both variables, as well as censored data (upper or lower limits). The \texttt{bandfit}
analysis for the $L_{\rm NVSS}$--$L_{\rm X,AGN}$ relation puts the chance for the null hypothesis of no correlation to less than $10^{-11}\%$ ($\sim8\sigma$). This correlation is not simply a consequence of plotting distance vs. distance; it is equally significant if the distance dependencies of both parameters are removed. The best fit correlation can be described as
\begin{equation}
\log L_{\rm X,AGN}=0.62 (\pm 0.10) \log \LNVSS + 19.11 (\pm 0.12).
\end{equation}

Similar correlations have been found in the past for AGN \citep[e.g.][and references therein]{HardcastleLxLr, BrinkmannLxLr}, with slopes ranging from $\sim 0.5$ for radio-loud quasars to $\gtrsim 1$ for radio-quiet quasars and radio galaxies. One must remember, however, that these results were obtained using total X-ray luminosities, not nuclear lumninosities as in our case. Also, previous samples have consisted mostly of objects in which the AGN dominate the X-ray emission, in contrast to our sample where the AGN is outweighed by the gas and LMXB contributions. If we replace $L_{\rm X,AGN}$ with the total gas luminosity
$L_{\rm X,gas}$, we find a steeper slope, similar to previous results for radio galaxies.

\cite{HardcastleLxLr} have argued that the existence of an $L_x$--$L_{\rm radio}$ correlation indicates that both X-ray and radio emission originate in a strongly Doppler boosted jet. If this were the case, then measured X-ray or radio fluxes might reflect primarily the orientation of the jet relative to the line of sight, rather than the intrinsic power of the AGN. In this picture, the shallower slope of the correlation for quasars could be explainable by a combination of beamed and isotropic X-ray emission, implying that X-rays would, at least in part, measure AGN power. Whether the shallow slope we find in the $L_{\rm NVSS}$--$L_{\rm X,AGN}$ relation indicates a similarity with the quasars is unclear. It is possible that our slope could be affected at the low-luminosity end by occasional misidentification of individual LMXBs in the galaxy centers as AGN. However, if $L_{\rm NVSS}$ or $L_{\rm X,AGN}$ were only reflecting the jet orientation, we would not expect either to be correlated with gas morphology, as we do see.

A survey of radio emission from normal elliptical galaxies by \citet{WrobelRadioorigin} suggests that the majority of their radio flux is due to AGN activity rather than star formation. The fact that nuclear X-ray and radio luminosities show a tight relationship supports this conclusion for our sample, as well as indicating that we have properly identified and isolated the central AGN in the X-ray images. We conclude that both the radio and the X-ray AGN luminosities are reliable tracers of the central AGN activity in our sample, most likely due to jets powered by gas accreting onto a supermassive black hole.

\subsection{The Galaxy Environment} \label{s4.environment}

To quantify the likelihood of interactions with neighboring galaxies, we measure the projected galaxy density in the vicinity of our sample galaxies by counting the number of neighbors listed in the 2MASS extended source catalog \citep{2MASS}. We identify all galaxies within a projected radius of 100~kpc at the distance of the target as ``neighbors.'' We restrict this distance to a maximum angular size of 15~arcmin, to minimize contamination by random foreground or background objects. Despite the smaller physical extraction radius for nearby objects, the number of neighbors is still sufficient to
derive a galaxy density, as the incompleteness limit drops significantly, and lower-luminosity neighbors can be identified. Because the 2MASS catalog is flux-limited, we have to cope with a distance-dependent completeness limit. We use the $K$-band Schechter luminosity function from \citet{KochanekKSchechter} to correct for incompleteness. We integrate the luminosity function down to the completeness limit at the object's distance and divide by the integral of the luminosity function down to a reference luminosity, which we arbitrarily chose as $M_{\rm K}=-11.5$. This way we can calculate the fraction of all galaxies above the reference luminosity that 2MASS is able to detect. We find that, after correcting for incompleteness, our galaxy densities \rhomass\ are independent of distance. We list the values for $\rhomass$ in Table \ref{t.radiogalaxyprop} in units of galaxies per $\mpc^2$. The errors include only statistical errors due to the galaxy counting statistics; deriving systematic errors due to background objects or uncertainties in the galaxy luminosity function are beyond the scope of this paper.

Ideally, one would also like to restrict the list of neighbors in velocity space, to remove galaxies that are simply aligned along the line of sight. Such a three-dimensional restriction requires an additional unbiased source of radial velocities. However, available radial velocities are heavily biased toward ``more interesting'' regions of the sky and better studied objects. Thus, we refrain from using these velocities and use the projected galaxy density $\rhomass$ instead, a measure similar to the Tully density parameter \citep{TullyRho}, but well-defined for our entire sample. All conclusions based on \rhomass\ are reproducible with the Tully parameter instead.

The hot ISM morphology may also be affected by hydrodynamic interactions with a surrounding intra-cluster medium (ICM) or intra-group medium (IGM). One expects a transition to an IGM or ICM at large radii to be accompanied by an outward rise in gas temperature,  reflecting the higher virial temperature of the group or cluster. To assess the presence of a hotter ambient medium, we compute radial temperature profiles for our sample. The X-ray counts image of each galaxy is divided into elliptical annuli, according to the X-ray ellipticity profiles computed in Paper I, except for those with insufficient signal to fit ellipses, for which we revert to circular annuli. We then extract a source and background spectrum for each annulus and fit them with a two-component model in the CIAO analysis package Sherpa. The first component consists of an APEC \footnote{Astrophysical Plasma Emission Code} plasma model to represent the hot gas emission. A quantitative comparison with its better known predecessor, the Mekal model, shows nearly identical results. We fix the gas metallicity at the solar abundance value. Unresolved point sources are represented by a power-law model with the power law index fixed at 1.6. This ``universal'' spectral model is an adequate representation of the emission of low-luminosity low-mass X-ray binaries, as demonstrated in Paper I and determined independently by \citet{Irwin03}. We also add a multiplicative absorption component, for which we fix the hydrogen column density to the Galactic value, evaluated at the target position with the CIAO tool {\it Colden}\footnote{http://cxc.harvard.edu/toolkit/colden.jsp}. We repeat our spectral analysis for a few objects with the gas abundance as a free parameter, and find that our choice to fix them to the solar value does not affect the fitted temperature. Since the metallicity is poorly constrained by the fits in low signal-to-noise systems, we fix the metallicity for all of our galaxies, in order not to introduce systematic differences in the analysis. We then fit the temperature profile by a power law, and derive a mean logarithmic temperature gradient between 2 and 4 J-band effective radii ($R_J$): $\alpha_{24}\equiv d \ln T /d \ln R |_{2-4\,R_J}$. Values of $\alpha_{24}$ are listed in the last column of Table \ref{t.asymgalaxyprop}.

One can consider the galaxy density $\rhomass$ as an indicator of the likelihood of gravitational galaxy interactions through close encounters or mergers, similar to the Tully galaxy density parameter. The outer temperature gradient, on the other hand, is a measure of the presence of hot ambient gas and thus a proxy for the likelihood of hydrodynamic interactions with this gas. A much more thorough analysis of the temperature profiles present in our sample will be the main focus of Paper III.


\section{Results} \label{s4.results}

\subsection{X-ray Ellipticity and Asymmetry}\label{s4.asymeps}

\begin{figure}
\begin{center}
\includegraphics[width=0.45\textwidth]{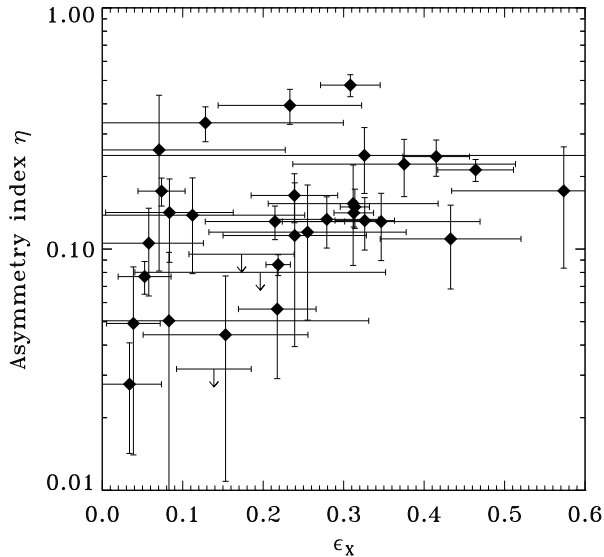}
\end{center}
\caption{The X-ray asymmetry index $\eta$ as a function of X-ray
ellipticity $\epsilon_{\rm X}$. The observed positive trend
suggests a common underlying cause for both the large measured X-ray
ellipticities and the asymmetries.\label{f.asymeps}}
\end{figure}

Figure \ref{f.asymeps} shows a comparison between the mean X-ray ellipticities between $0.6-0.9$ optical radii and the asymmetry index
$\eta$. There is a trend for flatter galaxies to be also more asymmetric, with a \texttt{bandfit} correlation strength of 99.72\%. There is a clear deficit of galaxies with flattened X-ray emission, but small asymmetry, which suggests a common underlying cause for both. The ellipticity may best be interpreted as simply a mean measure of the $m=2$ Fourier amplitude of the departure from circular symmetry. This is consistent with our results from Paper I, where we find that the gas ellipticities are uncorrelated with the ellipticities of the starlight, and thus are not simply determined by the shape of the smooth underlying gravitational potential.

\subsubsection{AGN Influence on Gas Morphology}\label{s4.asymagn}

\begin{figure*}
\begin{center}
\includegraphics[width=0.9\textwidth]{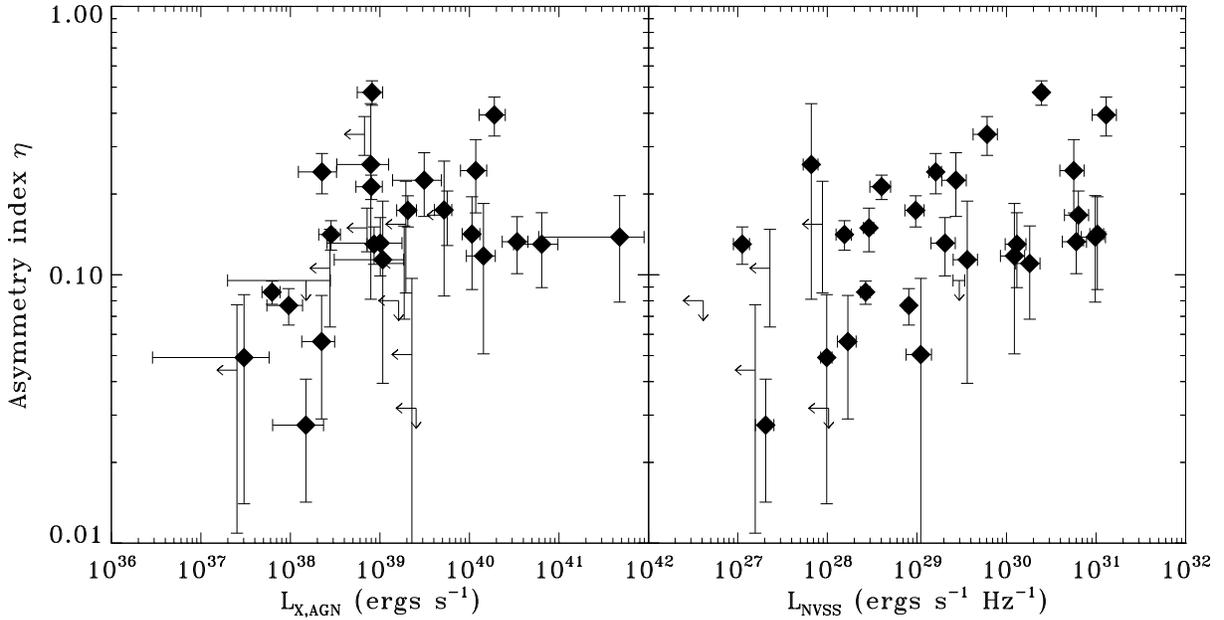}
\end{center}
\caption{Asymmetry index $\eta$ as a function of X-ray AGN luminosity
$L_{\rm X,AGN}$ (left) and $20\cm$ radio continuum power $L_{\rm
NVSS}$ taken from NVSS (right). Both plots indicate that the central
AGN luminosity is positively correlated with the observed asymmetry in
the gas, with the correlation extending all the way down to the
weakest AGN. \label{f.AGNasym}}
\end{figure*}

Figure \ref{f.AGNasym} shows the dependence of the asymmetry index $\eta$ on the two independent measures of AGN activity: the X-ray luminosity $L_{\rm X,AGN}$ of the central point source (left panel) and the NVSS radio power $L_{\rm NVSS}$, integrated over three optical radii (right panel). Both AGN properties are correlated with $\eta$, in the sense that galaxies hosting stronger AGN have stronger morphological asymmetries in their hot ISM. Analysis with \texttt{bandfit} rejects the null hypothesis of no correlation between $L_{\rm NVSS}$ and $\eta$  at the 98.7\% confidence level (see Table \ref{t.significances}). The same analysis for $L_{\rm X,AGN}$ and $\eta$ yields a slightly lower confidence value of 97.1\%. If we regard $L_{\rm X,AGN}$ and $L_{\rm NVSS}$ as independent measures of the same phenomenon (\S\ref{s4.agnradioxray}), we can compute the joint probability of the null hypothesis of no correlation between general AGN activity and asymmetry, which is then rejected at the 99.96\% level. Similar results are obtained from bootstrapping and from a Spearman rank analysis ignoring the upper limits. This AGN--asymmetry connection is strong evidence that the central AGN is {\it generally} responsible for disturbing the hot gas morphology even in normal elliptical galaxies and provides further support for the importance of AGN feedback.

A comparison with our simulations in \S\ref{s4.asymmetrytest} suggests two possible interpretations for the increase of asymmetry with AGN luminosity: the number of inflated bubbles, or the strength of the created depressions. The asymmetry index is equally sensitive to both and cannot distinguish between them. However, a qualitative investigation of the gas morphologies (see Paper I) suggests that depression strengths are more important. The examples in Figure \ref{f.asymexample} already show that galaxies with larger $\eta$ values tend to have stronger surface brightness depressions, possibly caused by the stronger interaction of more powerful radio sources with the ISM.

\begin{deluxetable}{lrr}
\tablewidth{0pt}
\tablecaption{Significances for correlations with $\eta$. \label{t.significances}}
\tablehead{
\colhead{Second Variable} & \colhead{$P_{Null}$}}
\startdata
$\log L_{\rm NVSS}$ \& $\log L_{\rm X,AGN}$ (jointly)  & 0.04\% \\
$\alpha_{24}$         & 0.2\% \\
$\log L_{\rm NVSS}$   &  1.3\% \\
$\log L_{\rm X,AGN}$ & 2.9\% \\
$\log \rhomass$             & 24.9\% \\
$\log \rho_{\rm Tully}$  & 45.1\% \\
\enddata
\end{deluxetable}

\subsection{Environmental Influence on Gas Morphology}

\begin{figure}
\begin{center}
\includegraphics[width=0.45\textwidth]{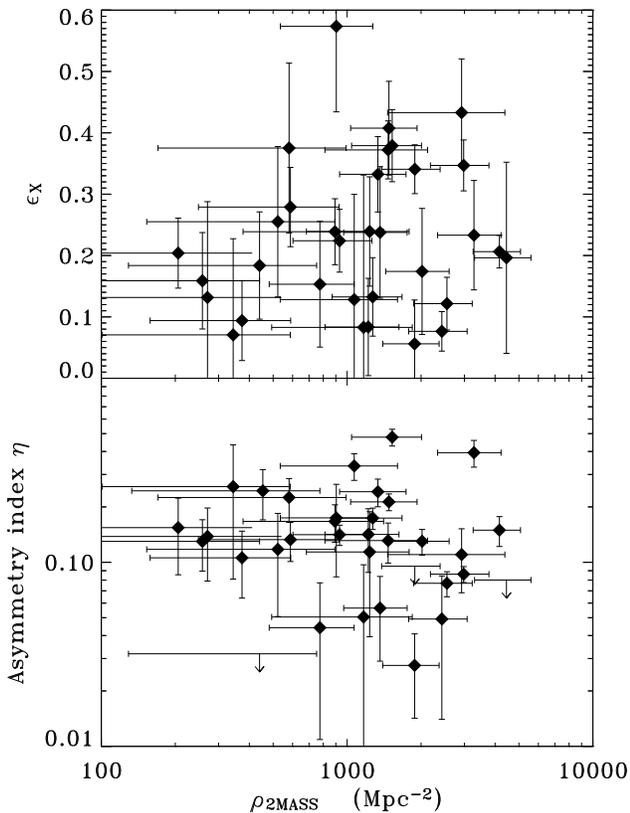}
\end{center}
\caption{X-ray ellipticity $\epsilon_{\rm X}$ (top) and asymmetry index $\eta$ (bottom) as a function of projected galaxy density
$\rhomass$. Both plots indicate that environment is not the driving factor causing the ellipticities and asymmetries. \label{f.environment}}
\end{figure}

The probability of interactions with neighboring galaxies through mergers, tidal interactions or close encounters naturally increases with the local number density of galaxies. Thus, if interactions are important in influencing the gas morphology of elliptical galaxies, we would expect a correlation between morphological parameters and the projected galaxy density $\rhomass$. However, Figure \ref{f.environment} shows that neither the gas ellipticity (upper panel), nor the asymmetry index (lower panel) shows any trend with this measure of environment. A \texttt{bandfit} analysis yields no statistically significant correlation in either case (Table \ref{t.significances}). Bootstrapping and a Spearman rank analysis give consistent results.

To ensure that our conclusions are independent of our definition of galaxy density, we repeat the analysis with other environmental measures of galaxy density: the Tully density parameter $\rho_{\rm Tully}$ \citep{TullyRho}, the distance to the 10th closest neighbor in the 2MASS extended source catalog \citep{2MASS}, and the number of galaxies associated with a galaxy group, as listed in the Lyon Group of Galaxies \citep[LGG;][]{LGG} catalog. All of these measures of environment yield the same result: the galaxy density is not correlated with the observed asymmetries or large ellipticities of the hot gas. Thus, the X-ray gas morphology is not mainly driven by interactions with companions, although they may still be important for some individual objects.

\begin{figure}
\begin{center}
\includegraphics[width=0.45 \textwidth]{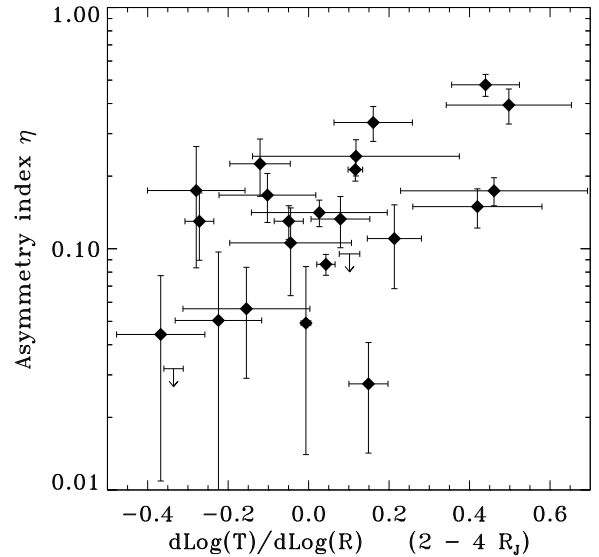}
\end{center}
\caption{Asymmetry index $\eta$ as a function of the outer temperature gradient $\alpha_{24}$, an indicator for the presence of hotter ambient gas. The correlation is such that galaxies in a hot gas environment are more asymmetric. Note that the sample includes both rising ($\alpha_{24}>0$, hotter ambient gas present) and falling ($\alpha_{24}<0$, no ambient gas present) temperature profiles. \label{f.tslope24asym}}
\end{figure}

In the absence of galaxy-galaxy interactions, hydrodynamic interaction with a hot ambient medium is still possible, through ram pressure, shocks, pressure confinement or hydrodynamic instabilities at the contact interface. Two galaxies in our sample are already known to be affected by ram-pressure: NGC~4472 \citep{BillerNGC4472} and NGC~1404 \citep{Machacek}. However, in both cases, the radius within which we compute the asymmetry index is smaller than the radius at which the interaction with the ambient medium becomes obvious. To assess the sample in general, we use the outer temperature gradient $\alpha_{24}$ as a proxy for a hot ICM, IGM, or circumgalactic gas. A larger gradient suggests a larger ambient gas pressure. Figure \ref{f.tslope24asym} shows that asymmetry indeed increases as a function of $\alpha_{24}$. A \texttt{bandfit} analysis rejects the null hypothesis at the 99.8\% level (Table \ref{t.significances}), with similar results obtained from bootstrapping and Spearman rank correlations.

\section{Discussion} \label{s4.discussion}

\subsection{Causes of Disturbances in the Gas Morphology}

Our comparison of optical and X-ray morphologies in Paper I revealed that even well within one effective radius, where the gravitational potential is dominated by the stellar component \citep{MamonDarkmatterI, HumphreyDarkmatter}, there is no correlation between stellar and gas ellipticities, very much at odds with predictions from hydrostatic equilibrium. In fact, many X-ray ellipticities even exceed those of the starlight, quite the opposite of what is expected from purely gravitational effects. We also noted in Paper I the prevalence of asymmetries in the hot gas, which we have now statistically quantified in this paper with the asymmetry index $\eta$. We find that the large observed ellipticities and the amount of asymmetry in the gas are correlated, pointing toward a common cause that dictates the overall X-ray morphology.

\begin{figure*}
\begin{center}
\includegraphics[width=0.45\textwidth]{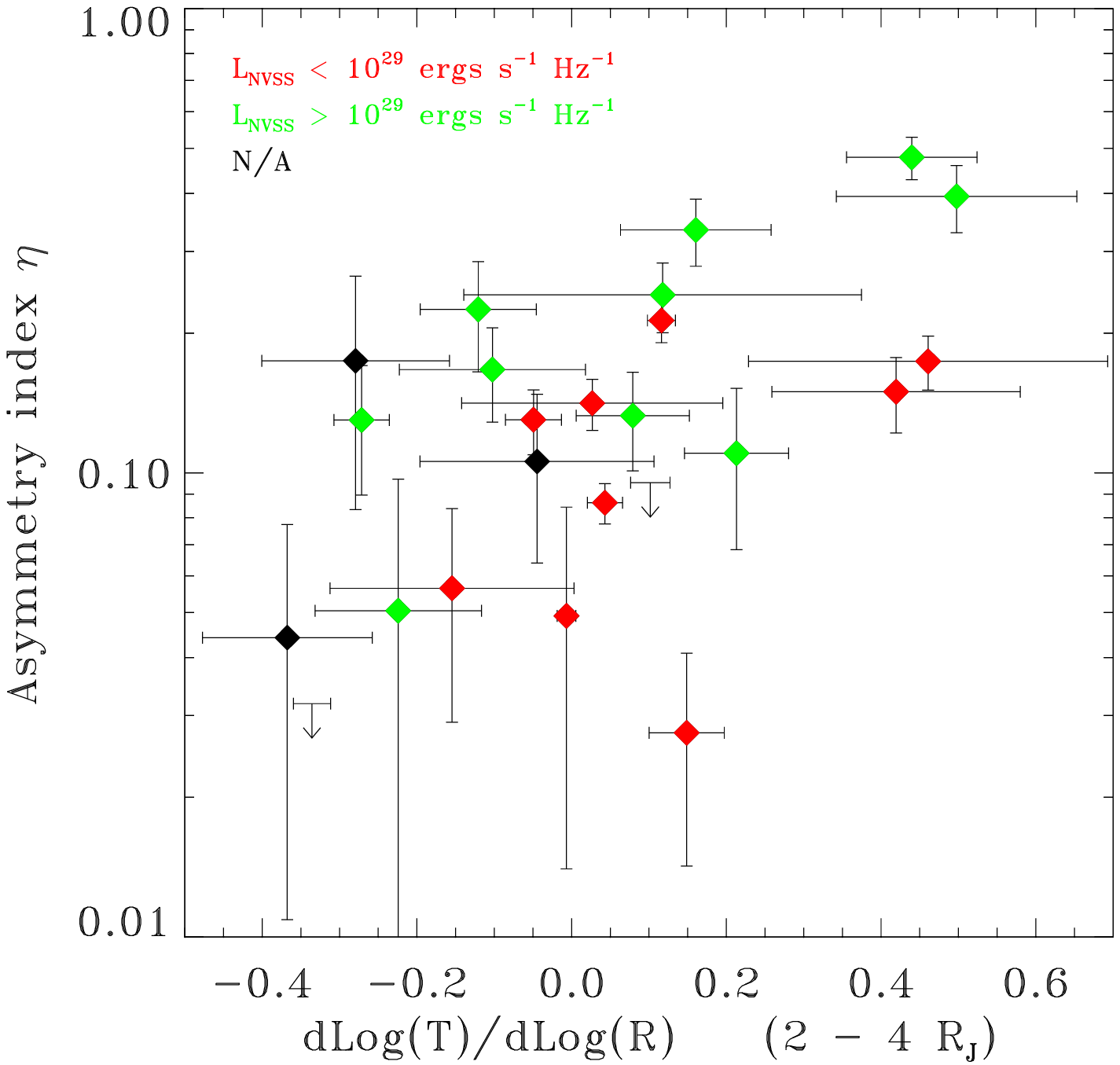}
\includegraphics[width=0.45\textwidth]{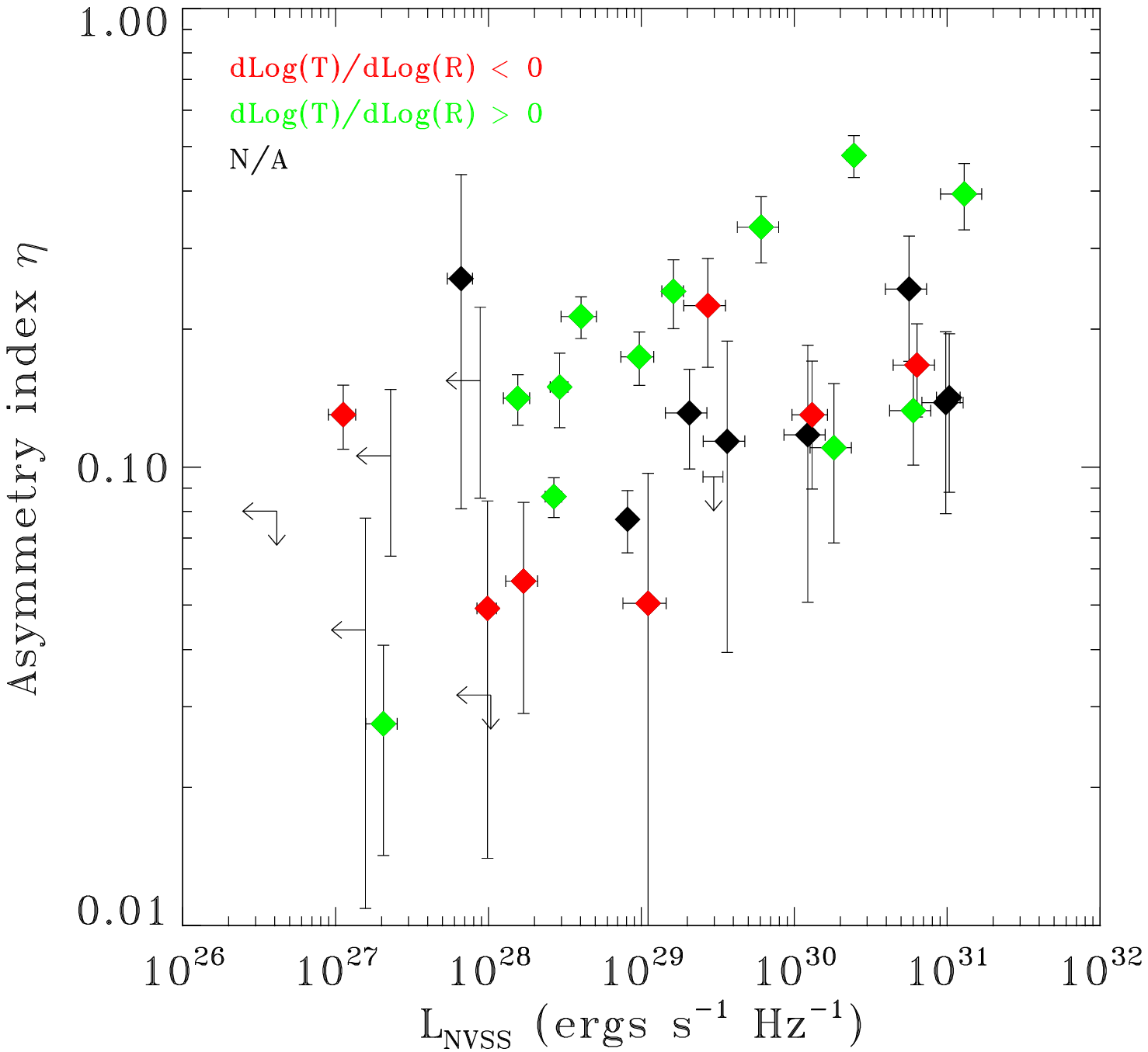}
\end{center}
\caption{Similar to Figures \ref{f.tslope24asym} and \ref{f.AGNasym}. {\em Left\/}: correlation between $\eta$ and the outer temperature gradient $\alpha_{24}$. Red colors indicate low-luminosity ($\LNVSS < 10^{32}\erg\s^{-1}\,{\rm Hz}^{-1}$), and green colors higher luminosity ($\LNVSS >10^{32}\erg\s^{-1}\,{\rm Hz}^{-1}$) AGN. {\em Right\/}: correlation between $\eta$ and radio luminosity $L_{\rm NVSS}$. Red points indicate a negative, and green a positive, temperature gradient $\alpha_{24}$. In each case the residuals from the mean relation are correlated with the third variable. On can consider the more fundamental correlation to be a 2-d plane in $\log \LNVSS$--$\alpha_{24}$--$\log \eta$ space.
\label{f.coloredasym}}
\end{figure*}

In this Paper, we have shown that feedback from the central AGN (Figure \ref{f.AGNasym}) and hydrodynamic interactions with the ambient medium (Figure \ref{f.tslope24asym}) are the dominant causes of these asymmetries. Our analysis of the AGN-morphology correlation shows that this phenomenon is not restricted to powerful radio sources, but forms a continuous sequence down to the weakest AGN. The correlation with the outer temperature gradient $\alpha_{24}$, on the other hand, suggests that the presence of a hot ambient medium has a similarly strong effect on the gas morphology. Figure \ref{f.coloredasym} shows that both correlations together dictate the disturbance in the gas. In Figure \ref{f.coloredasym}a we have colored the residuals in the $\alpha_{24}-\eta$ correlation according to AGN luminosity, with red indicating low ($\LNVSS < 10^{32}\erg\s^{-1}\,{\rm Hz}^{-1}$), and green high, radio power. In Figure \ref{f.coloredasym}b we color the residuals in the $\LNVSS-\eta$ correlation according to $\alpha_{24}$, with red now indicating a negative and green a positive temperature gradient. (Black points indicate objects for which these values are unavailable.) In each case, the residuals are clearly correlated with the third quantity. Replacing \LNVSS\ with the X-ray AGN luminosities $L_{\rm X,AGN}$ yields comparable results. Note that the scatter about the mean correlations is about the same in each case, indicating that AGN and ambient medium have comparable effects on the gas morphology. 

\subsection{Implications for AGN Feedback in Elliptical Galaxies}

The great majority of galaxies in our sample are not conspicuous AGN; most harbor only weak, unresolved radio sources. Nevertheless, we find a correlation between hot gas asymmetries and the radio and X-ray AGN luminosities, down to the lowest luminosities detectable by NVSS, FIRST, or {\it Chandra.} These results solidify the AGN as an influential feedback mechanism for normal elliptical galaxies in general, even where they do not obviously harbor powerful nuclear sources.

Intermittent AGN activity may stir the hot gas, pushing it far enough out of equilibrium that information about the shape of the underlying potential is lost. That we observe morphological disturbances in nearly all elliptical galaxies, and that some objects even reveal signatures of multiple outbursts \citep[e.g.][]{OsullivanNGC4636}, suggest that the duty cycle of the AGN is shorter than or comparable to the sound crossing time. Thus, the hot gas is continually disturbed, without having time to resettle into equilibrium.

\begin{figure}
\begin{center}
\includegraphics[width=0.4\textwidth]{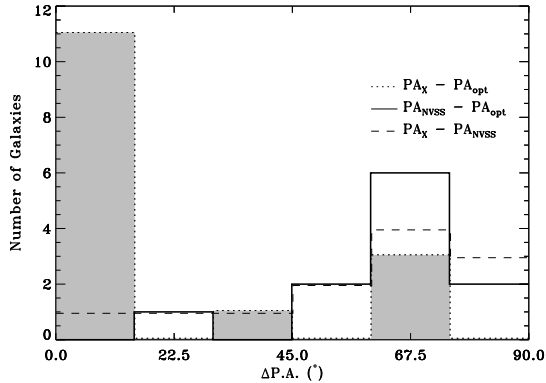}
\end{center}
\caption{Histogram of absolute differences between optical (${\rm PA}_{\rm opt}$), radio (NVSS 20cm continuum, ${\rm PA}_{\rm NVSS}$), and X-ray gas (${\rm PA}_{\rm X}$) position angles. The grey shaded histogram with dotted contours shows a tendency for galaxies to have X-ray and optical isophotes aligned. The solid line shows an anti-correlation between optical and radio position angles, and the dashed line a slight anti-correlation between radio and X-ray position angles, suggesting that the radio source is responsible for the X-ray--optical alignment. \label{f.histoparadio}}
\end{figure}

Despite the lack of correlation between optical and X-ray ellipticities (Paper I), we do note a weak tendency for the hot gas isophotal major axes to be aligned with those of the starlight. To quantify this trend, we extract a subsample of 15 galaxies that have X-ray ellipticities $\epsilon_X>0.2$ in order to have well-defined position angles (PA). The PA differences between the gas and the stellar component, in $15^\circ$-wide bins, are shown as the grey-shaded histogram in Figure \ref{f.histoparadio}. A Kolmogorov-Smirnov (KS) test puts the probability for a chance correlation at only 0.05\%. 

While this alignment may reflect the orientation of the underlying gravitational potential, it may also be the result of interactions with the central radio source. Our sample exhibits an anti-correlation between radio and optical major axes, which is shown as the solid histogram in Figure \ref{f.histoparadio}. This controversial phenomenon has been observed early on by \citet{PalimakaRadioopt}, but has not been confirmed in subsequent studies \citep{SansomRadioopt}. Nevertheless, a KS-test yields only a 1.1\% probability for the null hypothesis of uncorrelated orientations in our small sample of 11 galaxies with both available radio and optical major axes.

In case of a relatively stable radio jet axis, one would expect the jet to preferentially evacuate gas along the axis, creating a misalignment between radio and X-ray gas emission. We do observe a very slight anti-correlation, shown in the dashed histogram in Figure \ref{f.histoparadio}, but only 12 galaxies have both reliable X-ray and radio position angles. The large error bars on both radio and X-ray position angles make it difficult to ascertain the significance of the correlation, which a KS-test puts nominally at 90\%. Extracting the X-ray position angles at a larger radius increases the significance; but the detection of this misalignment is marginal and awaits confirmation with a larger sample and deeper radio data. With our current small sample, a two-sided KS test yields a 22.3\% probability that the radio--X-ray anticorrelation and the X-ray--optical alignment are drawn from the same distribution, which rises to 80.6\% at a larger extraction radius. Thus, the data are consistent with the alignment being caused by the effects of the radio source on the hot ISM. The competing effect of the interaction with the ambient medium may also work against a proper detection of this misalignment. However, the position angle may also represent the only information that is still inferable about the underlying gravitational potential, namely its orientation.

\subsection{Implications for Interactions with the Intergalactic Medium}

We find that the observed ellipticities and asymmetries in the hot gas are not mainly a consequence of interactions with neighboring galaxies. On the other hand, interactions with the ambient medium seem to be as important as the interplay with the central radio source. We expect the impact of an external ambient medium to increase with distance from the center. However, our analysis is confined to a region within {\it Chandra}'s limited field of view, restricted further by our need for sufficient signal to fit isophotes. Thus, our asymmetry index is insensitive to structure outside a radius which depends on the average surface brightness of the object and the exposure time of the observation. To verify that the $\alpha_{24}-\eta$ correlation is not simply a consequence of a variable extraction radius, we have checked for a correlation between this radius and $\eta$ and we find none.

The majority of X-ray halos associated with elliptical galaxies extend well beyond the outer radii in Table \ref{t.asymgalaxyprop}, as was already demonstrated with observations with the {\it Einstein} and {\it ROSAT} satellites \citep[e.g.][]{Forman1985,OSullivan}. Interactions with the ambient medium may very well have an even stronger influence on the gas structure at larger radii. At this point we are not yet able to distinguish the effects of AGN from those of environment based on morphology alone. Nevertheless it is curious to note that we do not observe a correlation between the X-ray ellipticity and the outer temperature gradient, as one might expect. Perhaps a different definition of statistical asymmetry that is sensitive to ``lopsidedness'' may be useful in this regard. 

Alternatively, the existence of a positive outer temperature gradient could also be interpreted as the presence of an external fuel reservoir for the central AGN. \citet{BestAGNenvironment} finds a correlation between radio-loud AGN activity and environment, in the sense that a larger fraction of AGN in richer environment are switched ``on'' compared to isolated galaxies. In this scenario, the correlation between the outer temperature gradient and the gas asymmetry would be a result of an increased duty cycle for AGN outbursts, rather than hydrodynamic interactions with the ambient medium. In addition, the circumgalactic gas reservoir also acts as a pressure confinement, which would prevent radio jets and inflated cavities from simply leaving the system and provides a background contrast to actually identify asymmetries at larger radii. 


\section{Conclusions} \label{s4.conclusions}

We have extended our earlier work from Paper I on normal elliptical galaxies, which showed that optical and X-ray gas ellipticities are uncorrelated in the regions where stellar mass dominates, contrary to what is expected from gas in hydrostatic equilibrium. Instead, the hot gas is generally disturbed. To elucidate the nature of these asymmetries,  we have introduced the asymmetry index $\eta$, which measures the statistical deviation from a symmetric surface brightness model. This allows a systematic study of asymmetry even in relatively low-luminosity objects, in which asymmetric features are barely resolved.

We find a strong correlation between the asymmetry index and two independent measures of AGN activity: the radio continuum power at $20\cm$ from NVSS \citep{NVSS} and the X-ray AGN luminosity extracted from {\it Chandra} data. The observed AGN--asymmetry correlation persists all the way down to the weakest AGN, where the NVSS survey reaches its detection
limit. This is quite surprising, since these objects generally lack extended jet signatures in their radio images and are mostly detected as weak central point sources, if at all. We also find the asymmetry index to be correlated with the outer temperature gradient $\alpha_{24}$, a proxy for the presence of a hot intragroup or intracluster medium. The strength of this correlation is comparable to that of the AGN--asymmetry correlation, indicating that hydrodynamic interaction with the ambient medium is comparably important to AGN in determining the X-ray morphology of normal ellipticals. However, we find no such correlation with the density of neighboring galaxies, and no evidence that galaxy-galaxy interactions play a significant role in shaping the hot ISM in these systems. Alternatively, the presence of an outer temperature gradient may also indicate the availability of a fuel reservoir for the central AGN. In this case, the increased amount of asymmetries in the X-ray gas could indicate a greater duty cycle for AGN outbursts in these objects. 

The emerging picture is consistent with the AGN persistently stirring up the interstellar medium through intermittent outbursts, and strengthens the case for the AGN to be at least partly responsible for offsetting cooling in normal elliptical galaxies. We will address the impact of the central AGN on temperature and entropy profiles in detail in Paper III of this series \citep{DiehlEntropy}.


\acknowledgments We thank the anonymous referee for an insightful referee report which helped improve the manuscript. We have made use of data products from the Two Micron All Sky Survey, which is a joint project of the University of Massachusetts and the Infrared Processing and Analysis Center/California Institute of Technology, funded by the National Aeronautics and Space Administration and the National Science Foundation. Support for this work was provided by the National Aeronautics and Space Administration (NASA) through Chandra Awards G01-2094X and AR3-4011X, issued by the {\em Chandra X-Ray Observatory Center}, which is operated by the Smithsonian Astrophysical Observatory for and on behalf of NASA under contract NAS8-39073, and by National Science Foundation grant AST0407152.

\bibliographystyle{apj}
\bibliography{allreferences}

\appendix
\section{Correlation analysis with bandfit}\label{s.bandfit}

The \texttt{bandfit} algorithm fits a model to two-dimensional $(x,y)$ data that are, or are suspected of being, linearly correlated. The model is a linear band, described by an orientation angle $\phi$ relative to the $x$ axis, a perpendicular offset from the origin $z_0$, and an intrinsic Gaussian perpendicular width $\sigma_{\rm int}$. The likelihood of any observed datum $(x_i,y_i)$ is a function of the perpendicular distance to the midline of the band, and is straightforwardly calculated by convolving the band with the error ellipse for the datum. The algorithm is thus similar to line fitting using perpendicular residual minimization, which is not exactly the same as classical linear regression.

\texttt{Bandfit} is written to handle data with Gaussian errors in either or both variables, including correlated errors. Censored data (upper or lower limits in either or both variables) are also easily handled, and contribute to the likelihood via an integral over the convolved band up to their location (i.e., the likelihood is proportional to the erf function evaluated at the censored point).

The model parameters are found by maximizing the likelihood using the well known simplex (``amoeba'') method. Errors in the fitted parameters are returned in the form of a covariance matrix.

To assess a correlation, the data are first shifted to zero mean to minimize the covariance between $\phi$ and $z_0$. The square root of the $\phi\phi$ element of the covariance matrix can then be interpreted as the standard error in $\phi$. Tests with simulated data sets show that integrating over this Gaussian provides accurate confidence limits on the non-zero slope of the correlation.

\texttt{Bandfit} is implemented in IDL, and will be made available through the IDL astro library after a thorough workout by volunteer beta-testers.

\end{document}